\documentclass[12pt,preprint]{aastex63}
\usepackage{graphicx}
\usepackage[]{xcolor}

\received{...}
\revised{...}
\accepted{...}
\submitjournal{ApJ}

\begin{document}

\title{The Energetics of the Central Engine in the Powerful Quasar, 3C298}
\correspondingauthor{Brian Punsly}
\email{brian.punsly@cox.net}
\author{Brian Punsly}
\affiliation{1415 Granvia Altamira, Palos Verdes Estates CA, USA 90274}
\affiliation{ICRANet, Piazza della Repubblica 10 Pescara 65100, Italy}
\affiliation{ICRA, Physics Department, University La Sapienza, Roma, Italy}

\author{Christian Groeneveld}
\affiliation{Leiden Observatory, Niels Bohrweg 2, 2333 CA Leiden}

\author[0000-0001-6717-7685]{Gary J. Hill}
\affiliation{McDonald Observatory, University of Texas at Austin, 2515 Speedway, Austin, TX 78712, USA}
\affiliation{Department of Astronomy, University of Texas at Austin, 2515 Speedway, Austin, TX 78712, USA}

\author{Paola Marziani}
\affiliation{INAF, Osservatorio Astronomico di Padova, Italy}

\author{Gregory R. Zeimann}
\affiliation{Hobby-Eberly Telescope, 32 Mt. Locke Rd., McDonald Observatory, TX 79734, USA}
\affiliation{McDonald Observatory, University of Texas at Austin, 2515 Speedway, Austin, TX 78712 USA}

\author{Donald P. Schneider}
\affiliation{Department of Astronomy, Pennsylvania State University, 525 Davey Lab, University Park, PA 16802, USA}
\affiliation{Institute for Gravitation and the Cosmos, Pennsylvania State University, University Park, PA 16802, USA}

\begin{abstract}
The compact steep spectrum radio source, 3C 298, (redshift of 1.44) has the largest 178 MHz luminosity in the 3CR (revised Third Cambridge Catalogue) catalog; its radio lobes are among the most luminous in the Universe. The plasma state of the radio lobes is modeled with the aid of interferometric radio observations (in particular, the new Low Frequency Array observation and archival MERLIN data) and archival single-station data. It is estimated that the long-term time-averaged jet power required to fill these lobes with leptonic plasma is $\overline{Q} \approx 1.28 \pm 0.51 \times 10^{47} \rm{erg}\,\rm{s}^{-1}$, rivaling the largest time averaged jet powers from any quasar. Supporting this notion of extraordinary jet power is a 0.5 keV -10 keV luminosity of $\approx 5.2 \times 10^{46} \rm{erg}\, \rm{s}^{-1}$, comparable to luminous blazars, yet there is no other indication of strong relativistic beaming. We combine two new high signal to noise optical spectroscopic observations from the Hobby-Eberly Telescope with archival Hubble Space Telescope, Two Micron Survey and Galaxy Evolutionary Explorer data to compute a bolometric luminosity from the accretion flow of $L_{\rm{bol}} \approx 1.55 \pm 0.15 \times 10^{47} \rm{erg} \,\rm{s}^{-1}$.  The ratio, $\overline{Q}/L_{\rm{bol}}\approx 1$, is the approximate upper limit for quasars. Characteristic of a large $\overline{Q}/L_{\rm{bol}}$, we find an extreme ultraviolet (EUV) spectrum that is very steep (the ``EUV deficit" of powerful radio quasars relative to radio quiet quasars) and this weak ionizing continuum is likely a contributing factor to the relatively small equivalent widths of the broad emission lines in this quasar.
\end{abstract}
\keywords{black hole physics --- galaxies: jets---galaxies: active
--- accretion, accretion disks}

\section{Introduction}
The quasar 3C 298, at a redshift of z = 1.44, has the highest 178 MHz luminosity, $L_{178}$, of any source in the 3CR (revised Third Cambridge) catalog \citep{ben62,sal08}. The small angular size of 2\farcs5 estimated in \citet{spe89} makes this a member of a subclass of radio sources known as a compact steep spectrum radio source (CSS).
The CSS sources are a particular class of small extragalactic radio sources with a size less than the galactic dimension and this can be taken as a working definition \citep{ode98}. They typically have synchrotron self-absorbed spectra, in which the spectral peak frequencies, $\nu_{\rm{peak}}\sim 100$ MHz, before turning over to the characteristic steep spectrum. This low frequency value of $\nu_{\rm{peak}}$ distinguishes them for other peaked radio sources, gigahertz peaked sources (GPS) and high frequency peaked sources (HFP) \citep{ode98,ori08}. They could be frustrated by the denser galactic environment, but in general, it is believed that most are in the early stages of an evolutionary sequence in which the CSS sources are younger versions of the larger radio sources, $> 50 \rm{kpc}$ \citep{ode98}. Traditional estimates of long term time average jet power, $\overline{Q}$, are based on radio sources that are much larger than the host galaxy \citep{wil99,bir08,cav10}. Thus, these estimates of $\overline{Q}$ are not valid in general for CSS sources \citep{bar96,wil99}. One of our primary goals is to determine if the large value of $L_{178}$ for 3C 298 is indicative of one of the most powerful jets in the Universe or if it is amplified by the dissipative interaction of the expanding radio source with the host galaxy.
\par 3C 298 is also known for having very strong star-burst regions and a conical ionized wind along the jet direction \citep{pod15,vay17}. Since the bulge luminosity is extremely low for the virial estimated central black hole mass, $M_{bh}$, it was concluded that strong negative feedback is occurring in a conical outflow early in the
gas-rich merger phase \citep{vay17}. This highly dynamical system motivates a detailed analysis of the central engine of this quasar with an extraordinarily powerful jet. In particular, we explore the UV continuum and emission lines with the aid of the upgraded Hobby-Eberly Telescope\footnote{The Hobby-Eberly Telescope is operated by McDonald Observatory on behalf of the University of Texas at Austin, Pennsylvania State University, Ludwig-Maximillians-Universit{\" a}t M{\" u}nchen, and Georg-August-Universit{\" a}t, G{\" o}ttingen. The HET is named in honor of its principal benefactors, William P. Hobby and Robert E. Eberly.} (HET, \citealt{ram94,hill21}) and archival Hubble Space Telescope (HST) observations in order to characterize the nuclear environment.

\par The paper is organized as follows. Section 2 compiles the radio data with a particular emphasis on the new Low Frequency Array (LOFAR) photometry and the International LOFAR Telescope (ILT) images at frequencies below 100 MHz. Sections 3 and 4 fit the radio data with parametric models of the radio lobes and the nucleus. Section 5 finds an infinite set of physical models that correspond to this fit. The solution that has equal internal energy in the two lobes, bilateral symmetry, is one in which both lobes are near minimum internal energy. For this solution, we compute the jet power. Section 6 describes the new optical observations and provides a detailed discussion of the broad emission lines (BELs). The optical spectra are combined with archival data in Section 7 in order to find the continuum spectral energy distribution (SED) of the thermal emission from the accretion flow and its bolometric luminosity. $L_{\rm{bol}}$. In Section 8, we make a connection between the weak BELs (relative to the UV continuum) found in Section 6 and the steep extreme ultraviolet spectrum found in Section 7. Finally, in Section 9, we have the insight gained in the previous sections to address how to best approach the virial mass estimate of $M_{bh}$ for the particular continuum SED of 3C 298.
Throughout this paper, we adopt the
following cosmological parameters: $H_{0}$=69.6 km s$^{-1}$Mpc$^{-1}$, $\Omega_{\Lambda}=0.714$ and $\Omega_{m}=0.286$ and use Ned Wright’s Javascript
Cosmology Calculator \footnote{http://www.astro.ucla.edu/~wright/CosmoCalc.html} \citep{wri06}.

\section{Radio Observations}
3C 298 is a famous bright radio source that has been extensively observed. We have collected radio data in Table 1. Most of the data are archival integrated flux density measurements from 26.3 MHz to 15 GHz. We also include archival interferometric component data at 327 MHz, 5 GHz, 8.4 GHz and 15 GHz that have insufficient u-v coverage and sensitivity to detect all of the lobe flux density, rendering their value as merely lower bounds \citep{fan02,dal21,lud98,aku95,van92}. The source is very small with a projected size on the sky plane of 1\farcs5 typically claimed \citep{aku95}. However full track 1.66 GHz MERLIN observations detected diffuse lobe flux and found that the overall size is actually 2".5 \citep{spe89}. The small size and steep spectrum requires extremely challenging interferometric observations in order to characterize the lobes. Until recently the MERLIN observation was the only interferometric observation that had sufficient sensitivity and resolution to characterize the lobes. We review the archival low frequency integrated flux density observations and discuss a new LOFAR image in detail in this section. The low frequency emission is the most difficult to observe, yet the most crucial data for this analysis.
\par Most of the data in Table 1 are survey results and we apply a 10\% uncertainty in spite of much more optimistic estimates in their original published form. In our experience, this approach is consistent with the repeatability of such measurements with different telescopes. There are two exceptions. For the GMRT 150 MHz survey data in the TIFR GMRT Sky Survey Alternative Data Release (TGSS ADR), the 10\% uncertainty used in TGSS ADR does not appear to be vetted well on a case by case basis and can be considerably larger for individual sources; 15\% uncertainty is a more prudent choice \citep{hur17}. The best case is a low resolution Very Large Array (VLA) observation that does not resolve out the diffuse lobe flux density (such as L-band observations). For this best case, the uncertainty in the flux density measurements is 5\% based on the VLA manual\footnote{located at https://science.nrao.edu/facilities/vla/docs/manuals/oss/performance/fdscale}, see also \citep{per13}. High frequency measurements are difficult to find because both ATCA and VLA will resolve out lobe flux density.

\begin{figure}
\includegraphics[width= 1\textwidth,angle =0]{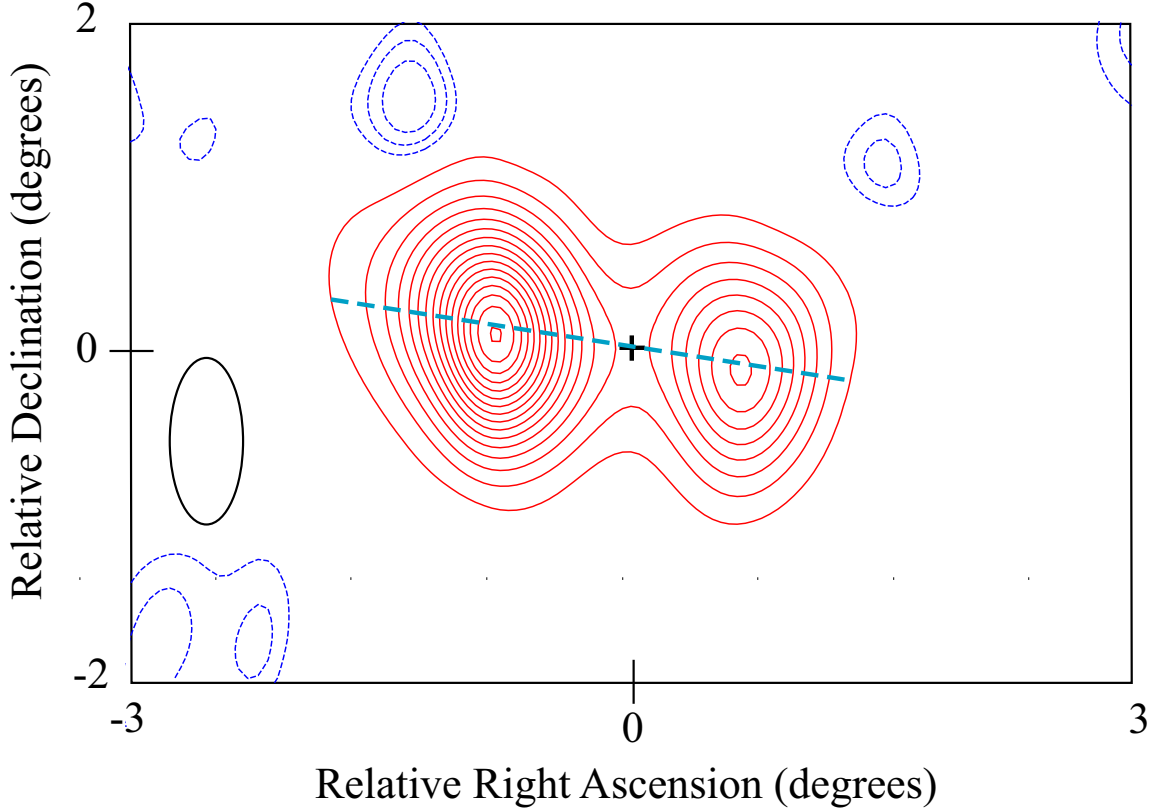}
\caption{\footnotesize{This wide bandwidth ILT image provides high sensitivity and is well-suited for the detection of diffuse lobe flux. The center frequency is 53.4 MHz and the bandwidth is 48.9 MHz. The red contour levels are in a linear scale of 2 Jy increments, 1.6Jy, 3.6 Jy, 5.6 Jy ,... 31.6Jy. The dashed-blue contour levels are -3$\sigma_{\rm{rms}}$, -4$\sigma_{\rm{rms}}$, -5$\sigma_{\rm{rms}}$ where $\sigma_{\rm{rms}}$ =0.22 Jy. The rms noise is driven primarily by calibration errors \citep{gro22}. The cross is the Gaia EDR3 optical position \citep{gai20}. This is the origin of the image. The restoring beam FWHM is represented by the ellipse in the lower left hand corner. The dashed light blue line is an approximate symmetry axis used to estimate arm length asymmetry in the source in Section 5.1.}}

\end{figure}

\begin{table}
    \caption{Radio Data for 3C298}
{\tiny\begin{tabular}{cccccc} \tableline\rule{0mm}{3mm}
$\nu_{\rm{obs}}$ & $\log{\nu}$ &Flux & Telescope & Absolute  & References\\
Observed  &  Rest   & Density &  & Flux Density &  \\
Frequency   & Frame  &   &  & Scale &  \\
(MHz)   & (Hz)  & (Jy) & & Factor &  \\
\tableline \rule{0mm}{3mm}
Integrated   & Flux  & Density & &  &  \\
26.3 & 7.81 & $54.59\pm 13.65$ & Clark Lake & 1.03   & \citet{vin75} \\
34.0 & 7.92 & $73.58\pm 21.00$ & Dutch Array of LOFAR & 0.98   & \citet{gro22} \\
38.0 & 7.97 & $84.68\pm 21.17$ & Large Cambridge Interferometer & 1.16   & \citet{kel69} \\
44.0 & 8.03 & $78.58\pm 22.00$ & Dutch Array of LOFAR & 0.98   & \citet{gro22} \\
53.5 & 8.12 & $98.10\pm 26.00$ & Dutch Array of LOFAR & 0.98   & \citet{gro22} \\
60.0 & 8.17 & $86.30\pm 9.97$ & LOFAR (AART-FAAC) & 1.00   & \citet{kui19} \\
63.0 & 8.19 & $108.89\pm 28.00$ & Dutch Array of LOFAR  & 0.98   & \citet{gro22} \\
73.0 & 8.25 & $101.04\pm 26.00$ & Dutch Array of LOFAR & 0.98   & \citet{gro22} \\
73.8 & 8.26 & $97.7\pm 9.77$ & VLA  & 1.00   & \citet{lan14} \\
150 &    8.56     & $59.59\pm 8.94$  & GMRT & 1.00  & \citet{int17,hur17} \\
160 &    8.59      & $63.20\pm 9.48$  & Culgoora Array & 1.00  & \citet{kuh81} \\
178 &    8.64      & $52.30\pm 7.85$  & Large Cambridge Interferometer & 1.00  & \citet{kuh81} \\
327  &   8.90     & $30.00\pm 3.00$  & VLA & 1.00  & VLA Calibrator List\tablenotemark{\tiny{a}}  \\
408 & 9.00 & $23.36\pm 2.34$  & Molonglo Telescope & 1.00 &  \citet{lar81} \\
1400 & 9.53 & $6.10\pm 0.31$  & VLA D-array & 1.00 & \citet{con98}  \\
1500 & 9.56 & $5.80\pm 0.29$  & VLA C and D-arrays & 1.00 & VLA Calibrator List\tablenotemark{\tiny{a}}  \\
1600 & 9.59 & $5.40\pm 0.27$  & VLA A-array & 1.00 &  \citet{aku95} \\
2640 & 9.81 & $2.89\pm 0.29$  & Effelsberg 100-m & 1.00 &  \citet{man09} \\
4850 & 10.07 & $1.57\pm 0.16$  & Parkes 64 m & 1.00 & \citet{gri95}  \\
8350 & 10.31 & $0.79\pm 0.08$  & Effelsberg 100-m & 1.00  &  \citet{man13} \\
8870 & 10.34 & $1.22\pm 0.12$  & Parkes 64 m & 1.00 & \citet{shi73}  \\
10070 & 10.42 & $1.03\pm 0.10$  & NRAO 140-ft & 1.00 & \citet{kel73}  \\
14900 & 10.56 & $0.69\pm 0.07$  & MPIfR 100-m & 1.00 & \citet{gen76}  \\
\tableline \rule{0mm}{3mm}
East Lobe &  &  &  &  &  \\
55.0 & 8.13 & $52.92\pm 13.23$  & International LOFAR & 0.98 & \citet{gro22}  \\
327 & 8.90 & $>8.55$ \tablenotemark{\tiny{b}} & Global VLBI & 1.00 & \citet{fan02,dal21} \\
1660 & 9.61 & $2.23\pm 0.22$  & MERLIN & 1.00 & \citet{spe89}  \\
4993 & 10.09 & $>0.51$\tablenotemark{\tiny{b}}  & MERLIN & 1.00 & \citet{lud98}  \\
8414 & 10.31 & $>0.28$\tablenotemark{\tiny{b}}  & VLA A-array & 1.00 & \citet{aku95}  \\
14965 & 10.56 & $>0.12$\tablenotemark{\tiny{b}}  & VLA A-array & 1.00 & \citet{van92}  \\
\tableline \rule{0mm}{3mm}
West Lobe &  &  &  &  &  \\
 55.0 & 8.13 & $22.54\pm 5.64$  & International LOFAR & 0.98 & \citet{gro22}  \\
327 & 8.90 & $>10.73$\tablenotemark{\tiny{b}}  & Global VLBI & 1.00 & \citet{fan02,dal21} \\
1660 & 9.61 & $2.00\pm 0.20$  & MERLIN & 1.00 & \citet{spe89}  \\
4993 & 10.09 & $>0.35$\tablenotemark{\tiny{b}}  & MERLIN & 1.00 & \citet{lud98}  \\
8414 & 10.31 & $>0.16$\tablenotemark{\tiny{b}}  & VLA A-array & 1.00 & \citet{aku95}  \\
14965 & 10.56 & $>0.05$\tablenotemark{\tiny{b}}  & VLA A-array & 1.00 & \citet{van92}  \\
\tableline \rule{0mm}{3mm}
Core + Jet &  &  &  &  &  \\
327 & 8.90 & $0.83\pm 0.12$  & Global VLBI & 1.00 & \citet{fan02,dal21} \\
1660 & 9.61 & $0.88\pm 0.09$  & MERLIN & 1.00 & \citet{spe89}  \\
4993 & 10.09 & $0.49\pm0.05$  & MERLIN & 1.00 & \citet{lud98}  \\
8414 & 10.31 & $0.43\pm0.04$  & VLA A-array & 1.00 & \citet{aku95}  \\
14965 & 10.56 & $0.27\pm0.04$  & VLA A-array & 1.00 & \citet{van92}  \\
22500 & 10.74 & $0.24\pm0.04$  & VLA A-array & 1.00 & \citet{van92}

\end{tabular}}
\tablenotetext{a}{\url{https://science.nrao.edu/facilities/vla/observing/callist}}
\tablenotetext{b}{A significant amount of lobe flux density is resolved out. Thus, this is a lower bound.}
\end{table}
\subsection{Low frequency Integrated Flux Densities} Low frequency observations were an important part of the early radio observations of extragalactic radio sources, including observations at frequencies $<100$ MHz. There are three major complicating issues, ionospheric propagation effects, the absolute flux density scale and the blending of sources in the low resolution telescopes. The ionospheric correction is the most deleterious. In particular, the varying index of refraction in both position and time creates a time varying shift and blurring of observed images and the associated phase delays wreak havoc on radio-interferometric observations \citep{deg18}. The first comprehensive study of 3C sources at low frequency was the 38 MHz data presented in \citep{kel69}. However, the absolute flux density scale was called into question in later studies due in large part to the drift in the brightness of the supernova remnant, Cas A, and the lack of well characterized flux density calibrators at low frequency \citep{rog73}. They determined that the absolute flux densities in \citet{kel69} were low by 18\%. A less well known 26.3 MHz study reanalyzed the \citet{rog73} conclusion using a deep survey with the Clark Lake telescope \citep{vin75}. The Clark Lake telescope had considerably higher resolution than the Penticon telescope ($\approx 0.5 \  \rm{deg}^{2}$ compared to $\approx 2 \ \rm{deg}^{2}$) used in \citep{rog69,rog73}. They found that the agreement with Penticon was excellent if only the sources with no confusion from nearby sources in the Penticon field were used. If the entire Penticon sample was used (including confused fields) the Penticon flux density scale was 5\% higher. Regardless, the Penticon flux density scale has been the most referenced calibration scale to this day and is used by the LOFAR team \citep{sca12}. Recently, this flux density scale was re-examined \citep{per17}. Examining their non-variable calibrators (neglecting 3C 380) suggests that the \citet{rog73} scale is 2\% high, including the case of 3C 196 which is the absolute flux density calibrator for 3C 298. The next to last column of Table 1 transforms all the archival data and the LOFAR data to the \citet{per17} absolute flux density scale (the scale factor). The only reliable $\sim 20$ MHz measurement of 3C 298 is the high resolution Clark Lake measurement \citep{vin75}. The radio emission from the cluster Abel 1890 gets blended into lower resolution fields used in other studies \citep{rog69}.
\par Note that the only scale factors different from 1.00 in Table 1 are at low frequency. The \citet{kel69} observations have the largest scale factor correction. The other low frequency scale factors differ only by a few percent from unity. The Large Cambridge Interferometer observations at 38 MHz are likely more uncertain than the 5\% - 15\% in \citet{kel69}, even after the scale correction factor is applied. The resolution of the telescope is insufficient to resolve 3.7 Jy of VLA Low-Frequency Sky Survey (VLSS) background sources at 74 MHz that were not de-blended \citep{lan14}. The spectral indices from 74 MHz to 38 MHz of the background sources are unknown, so there is no robust post-processing de-blending algorithm. We incorporate the uncertainty introduced by the inability to de-blend known background sources by conservatively choosing a large overall uncertainty of 25\%, similar to the LOFAR uncertainties in \citep{gro22}.

\subsection{LOFAR Observations}
\par To properly characterize the lobes, one requires observations at multiple low frequencies with high resolution in order to detect all the diffuse flux and resolve it into two distinct components. This was impossible until recently with the development of the International LOFAR Telescope (ILT), in particular the Low-Band Antenna, the LBA \citep{van13}. In 2020, an observation of 8 hours duration was performed in the LBA-OUTER configuration using the outer 48 of 96 antennae in each Dutch station. More details can be found in \citet{gro22}. Figure 1 is a remarkable previously unpublished high sensitivity LBA-OUTER image obtained at the band center of 53.4 MHz. The high sensitivity is achieved by using the full LBA-OUTER bandwidth of 48.9 MHz. It is this new image and the LOFAR data in \citet{gro22} that motivates our new detailed analysis of the radio lobes and confirms the 1.66 GHz MERLIN detection of a 2."5 linear size.
\par The Amsterdam ASTRON Radio Transient Facility And Analysis Centre (AART-FAAC), a parallel computational back-end to LOFAR, is the source of the 60 MHz data in Table 1. It processes data from six stations of the Dutch LBA-Outer configuration. It was concluded that the absolute flux density of these measurements is accurate to $10 \%$ \citep{kui19}. There are about 5 Jy of VLSS (74 MHz) background sources that are not resolved by AART-FAAC \citep{lan14}. Thus, the 60 MHz observation is chosen to have an uncertainty of 5 Jy added in quadrature with the $10\%$ estimate of \citet{kui19} in order to account for possible errors due to potential source blending. We included the flux density of each lobe separately from the 55 MHz ILT image in Table 1 \citep{gro22}. More details of the ILT observations can be found in \citep{gro22}. The uncertainty in the estimated ILT flux density of 3C\,298 in \citet{gro22} and Table 1 are largely calibration errors rather than errors inherent to the flux density scale.

\section{Synchrotron Self-Absorbed Homogeneous Plasmoids}
A high resolution 5 GHz MERLIN (0.05 mas synthesized beam) image reveals a detailed structure beyond the two lobes indicated in Figure 1 \citep{lud98}. There is a core that dominates with higher frequency very long baseline interferometry (VLBI) and is also prominent in the MERLIN image \citep{van92}. There is a jet emerging from the core towards the eastern lobe. The jet surface brightness vanishes as it propagates away from the core and then reveals itself again as it enters the eastern lobe. To even partially resolve the jet requires a resolution of $<200$ mas. Thus, observations that are capable of detecting the diffuse lobe flux are incapable of also detecting the jet and vice versa. We consider the source as consisting of three regions (or components). The western lobe, the core plus easterly directed jet, and the eastern lobe that includes the small piece of the jet as it enters the lobe. Our primary interest in determining the energetics of the radio source will be to estimate the energy stored in the luminous radio lobes. A simple homogeneous spherical volume model or plasmoid has historically provided an understanding of the
spectra and the time evolution of astrophysical radio sources \citep{van66}. Single zone spherical models are a standard technique even in blazar jet calculations out of practical necessity \citep{ghi10}. We have used this formalism to study a panoply of phenomena, major flares in a Galactic black
hole, a $\gamma$-ray burst and flares in a radio quiet quasar \citep{pun12,pun19,rey09,rey20}. Most importantly, we used this method in \citet{pun20} to study the radio lobes in the super-luminous CSS radio quasar, 3C 82 (which should be consulted for more details of the calculational method). The synchrotron self absorbed (SSA) turnover provides information on the size of the region that produces the preponderance of emission. This is useful because Figure 1 is resolution limited and provides no fine details of the emission regions. Furthermore, these models do not need to invoke equipartition in order to produce a solution.
\par These homogeneous models produce a simple to parameterize spectrum. A SSA power law for the observed flux density, $ S_{\nu_{\rm{obs}}}$, is the solution to the radiative transfer in a homogeneous medium such as a uniform spherical volume \citep{gin65,van66}:
\begin{eqnarray}
&& S_{\nu_{\mathrm{obs}}} = \frac{S_{\mathrm{o}}\nu_{\mathrm{obs}}^{-\alpha}}{\tau(\nu_{\mathrm{obs}})} \times \left(1 -e^{-\tau(\nu_{\mathrm{obs}})}\right)\;, \; \; \tau(\nu_{\mathrm{obs}})=\overline{\tau}\nu_{\mathrm{obs}}^{(-2.5 +\alpha)}\;,
\end{eqnarray}
where $\tau(\nu)$ is the SSA opacity, $S_{\mathrm{o}}$ is a normalization factor, $\overline{\tau}$ is a constant, and $\nu_{\rm{obs}}$ designates the observed frequency as opposed to $\nu$ which we use to designate the frequency in the plasma rest frame. The spectral index of the power law is defined in the optically thin region of the spectrum by $S_{\nu_{\mathrm{obs}}}= S_{\mathrm{o}}\nu_{\mathrm{obs}}^{-\alpha}$.
\begin{figure*}
\begin{center}
\includegraphics[width= 0.75\textwidth,angle =0]{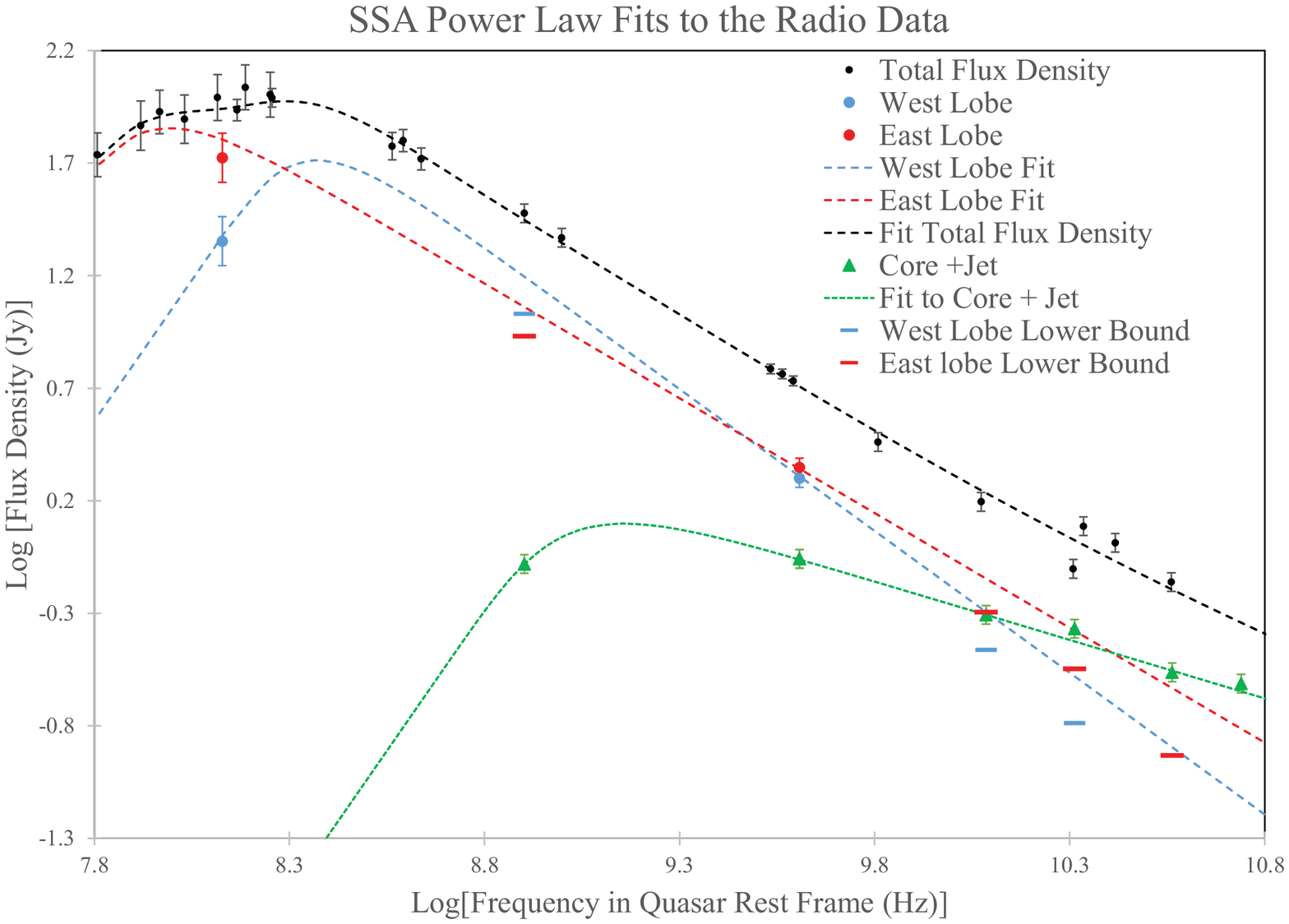}
\includegraphics[width= 0.75\textwidth,angle =0]{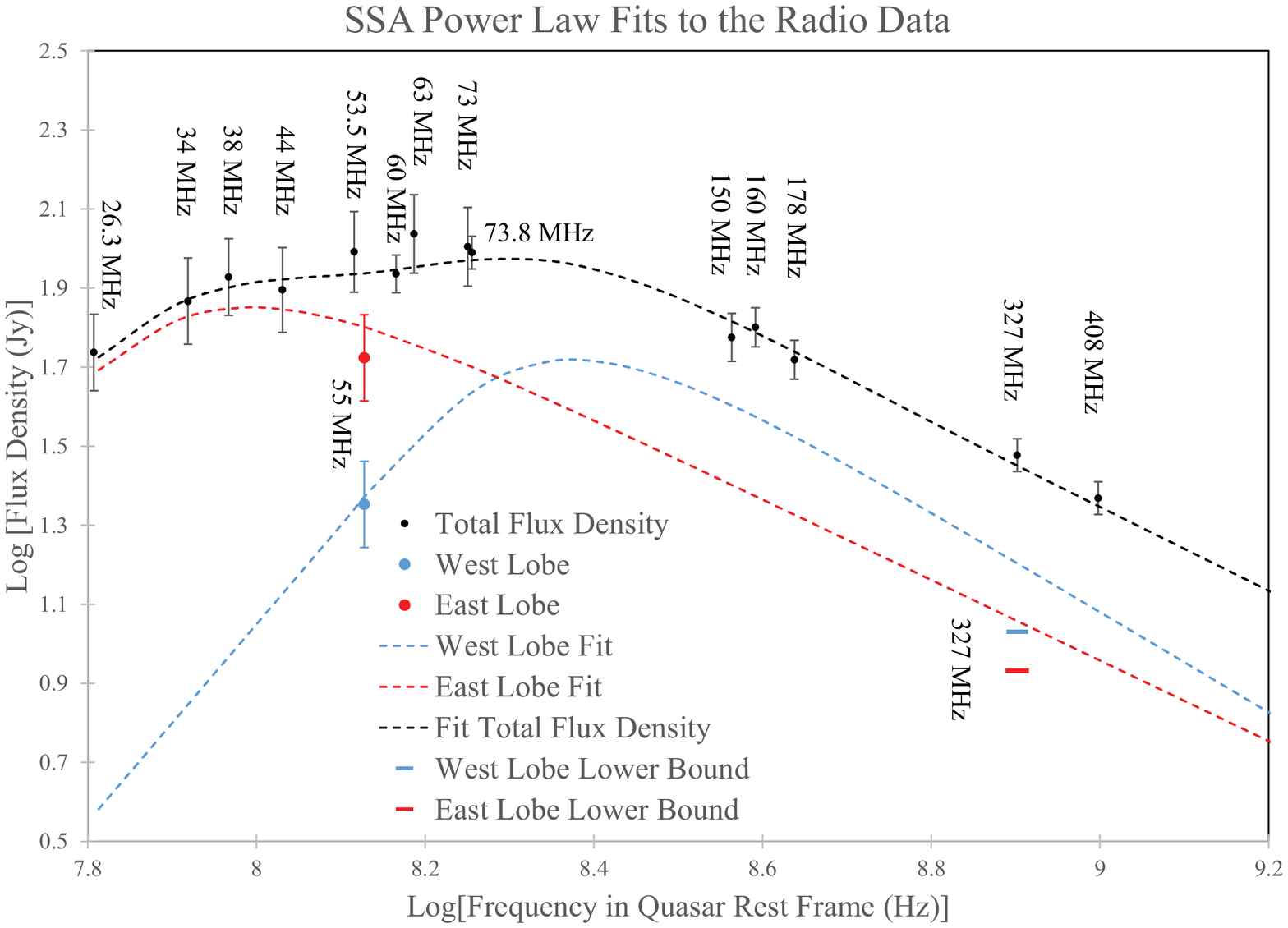}

\caption{The broad-band radio spectrum of 3C 298 is fit by three SSA power laws in the top panel. The radio data are from Table 1. Weighted residuals, as defined by Equation (2), are minimized for $\nu_{obs}\leq 4.85$ GHz (log[Frequency in Quasar Rest Frame] $\leq 10.1$ Hz). There are 22 total measurements that contribute to the sum in Equation (2). Four of these are the fitted lobe flux densities relative to the measured component flux densities. 18 are the sum of the fits (black-dashed curve) to the lobe flux densities and the core-jet flux density (which contributes minimally in this frequency range) relative to the integrated flux density measurements. The bottom panel is a magnified view of the densely sampled low frequency region. The observed frequencies of the data points are labeled in MHz in the bottom panel. The optically thin power law tail of the core, east lobe and west lobe have $\alpha=$ 0.52, 1.02 and 1.27, respectively in Equation (1) and $\nu_{\rm{peak}}$ is at 1.4 GHz, 95 MHz and 240 MHz in the quasar rest frame, respectively.}
\end{center}
\end{figure*}
\section{Fitting the Data with Three SSA Power Laws}
Thus motivated, we approximate the total spectrum by three SSA power laws, one for the western lobe, one for the eastern lobe, and one for the core plus jet. We determine the three SSA power laws that, in combination, minimize the residuals to the overall integrated spectral energy distribution. Since the uncertainty of the measurements in Table 1 are significant, we need to account for these in the calculation of the residuals of the fit to the data. It makes sense in curve fitting to assign the least amount of weight to points that are the least
reliable. This is properly accomplished statistically by weighting each point by the inverse square of its standard error \citep{ree89}. To incorporate this notion, we define a weighted residual, $\sigma_{\rm{res}}^{2}$,
\begin{equation}
\sigma_{\rm{res}}^{2}=\frac{1}{N}\sum_{i=1}^{N}
\left[\frac{(S_{i}-f_{i})^2}{\sigma_{i}^{2}}\right]\; ,
\end{equation}
where, ``i" labels one of the $N$ measured flux densities from Table 1, $f_{i}$ is the expected value of this flux density from the fitted curve, $S_{i}$ is the measured flux density and $\sigma_{i}$ is the uncertainty in this measurement. This quantity compares the fractional error of the fit to the data to that which is expected to occur just form measurement induced scatter. The smaller the value of $\sigma_{\rm{res}}^{2}$, the better that a particular fit agrees with the data.

\par We minimize the weighted residuals of the fit to the data, $\sigma_{\rm{res}}^{2}$, for 22 points in Table 1 (the 18 total flux density points with $\nu_{\rm{obs}} \leq 4.85$ GHz and the 4 flux densities for the lobes, LOFAR 55 MHz and MERLIN 1.66 GHz ). Since the total flux density is approximately the sum of the two lobe flux densities (the core-jet contribution is small in Figure 2 at $\nu_{obs}\leq 4.85$ GHz), the 18 $\nu_{\rm{obs}} \leq 4.85$ GHz total flux densities provide significantly more constraints on the lobe spectra than if we merely had a limited number of flux densities measured per lobe. The lower bounds are not used in the minimization of the residuals and are sufficiently low that they would not provide meaningful additional constraints. The spectral index of the optically thin power laws of the lobes fit the total flux density data tightly between 327 MHz and 1.6 GHz. The cutoff of data suitable for fitting at $\nu_{\rm{obs}} \leq 4.85$ GHz is motivated by the large core-jet contribution to the total flux density at higher frequencies that masks the pure lobe contribution. Furthermore, the core-jet contribution appears to create some scatter to a smooth monotonic fit as evidenced by Table 1 and Figure 2. This might represent intrinsic variability (compare the integrated flux density entry for 8350 MHz with that of 8870 MHz in Table 1), but the data is not of sufficiently quality to make such a claim at this point in time \footnote{The best way to determine this might be with a few years of monitoring with JVLA in D-array (so as not to resolve out flux) at X-band and U-band.}. Our primary scientific goal is to fit the radio lobes, so total flux density measurements that are strongly influenced by potential core variability are of little value to this process. In summary, there are 3 parameters for the SSA power law of each lobe. We are using 22 observational data points to constrain 6 total parameters by minimizing the residuals in Equation (2). The best fit is displayed in Figure 2. Note that the SSA turnover of the east lobe is real; an extrapolation of the power law seen at higher frequencies would grossly grossly over produce the flux density at 26.3 MHz and 34 MHz (see the bottom panel of Figure 2).

\begin{figure*}
\begin{center}
\includegraphics[width= 0.45\textwidth,angle =0]{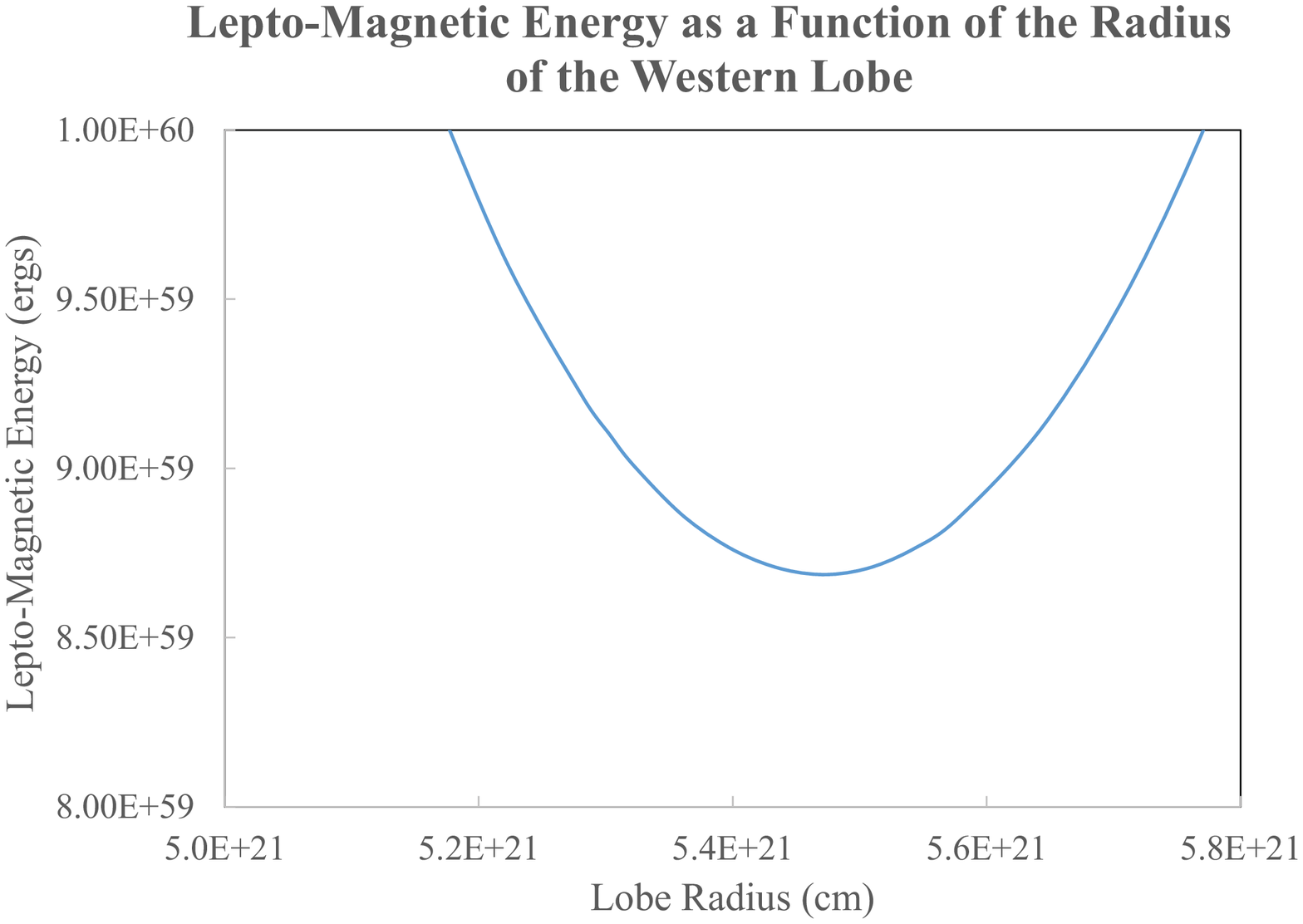}
\includegraphics[width= 0.45\textwidth,angle =0]{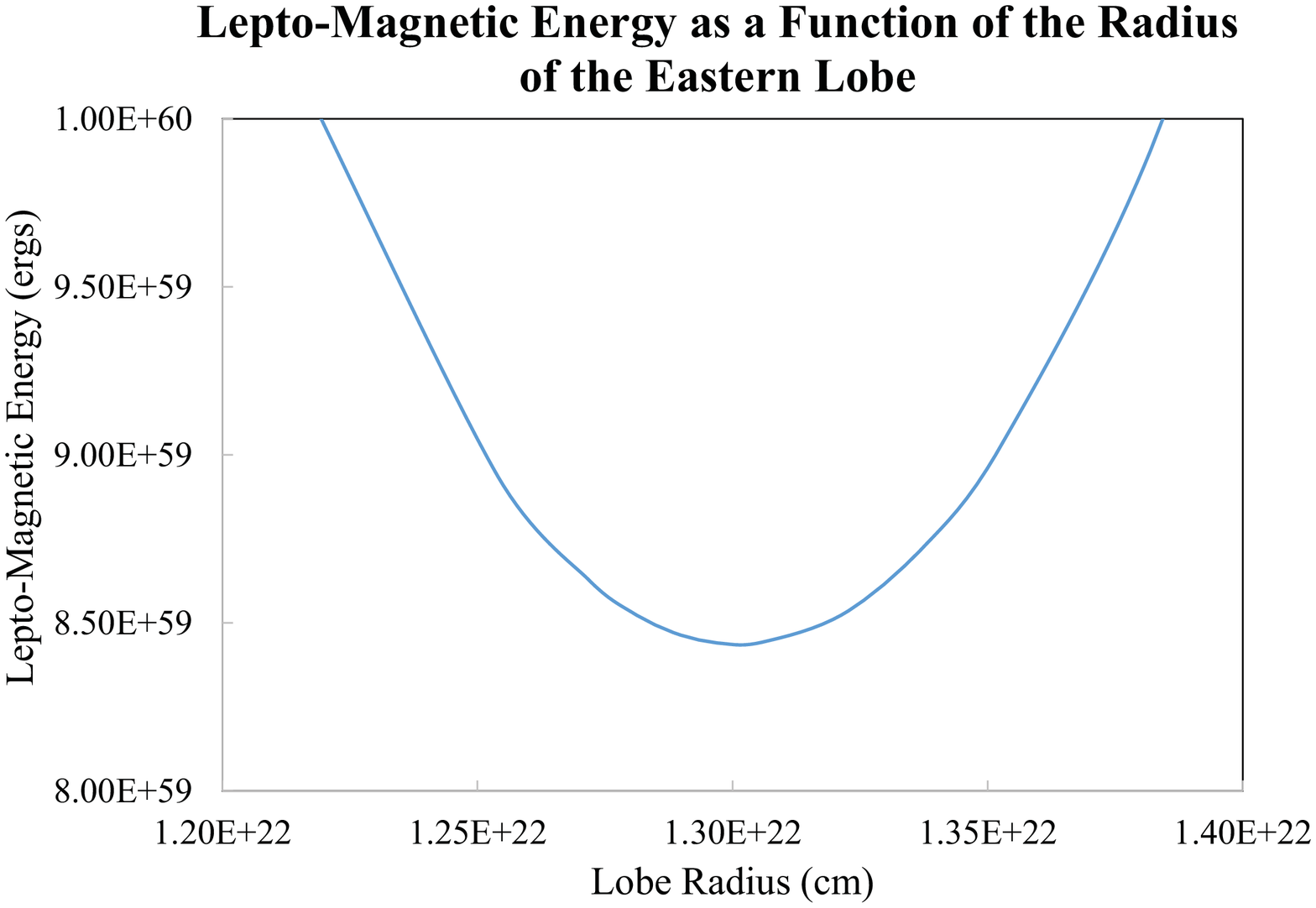}
\includegraphics[width= 0.45\textwidth,angle =0]{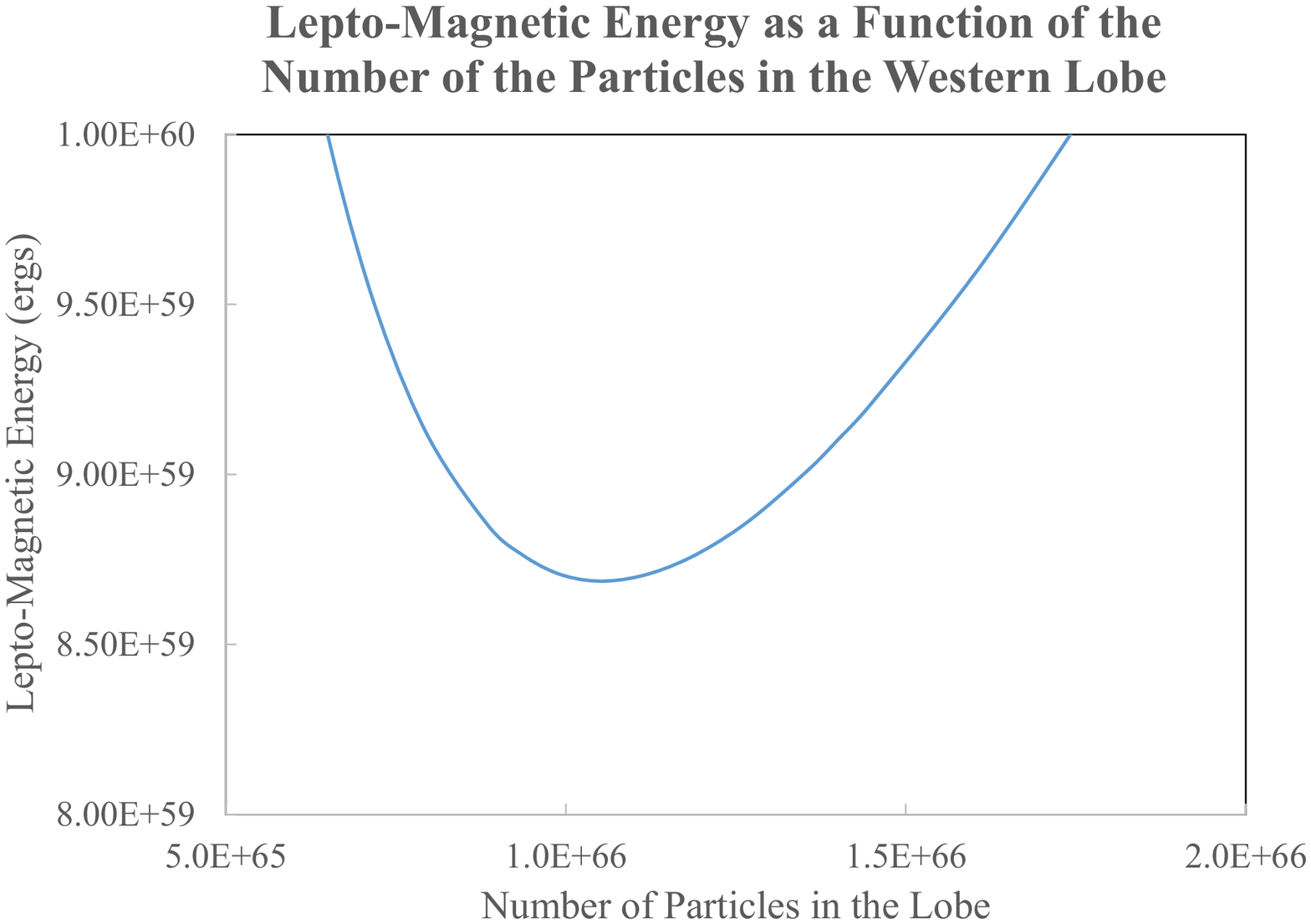}
\includegraphics[width= 0.45\textwidth,angle =0]{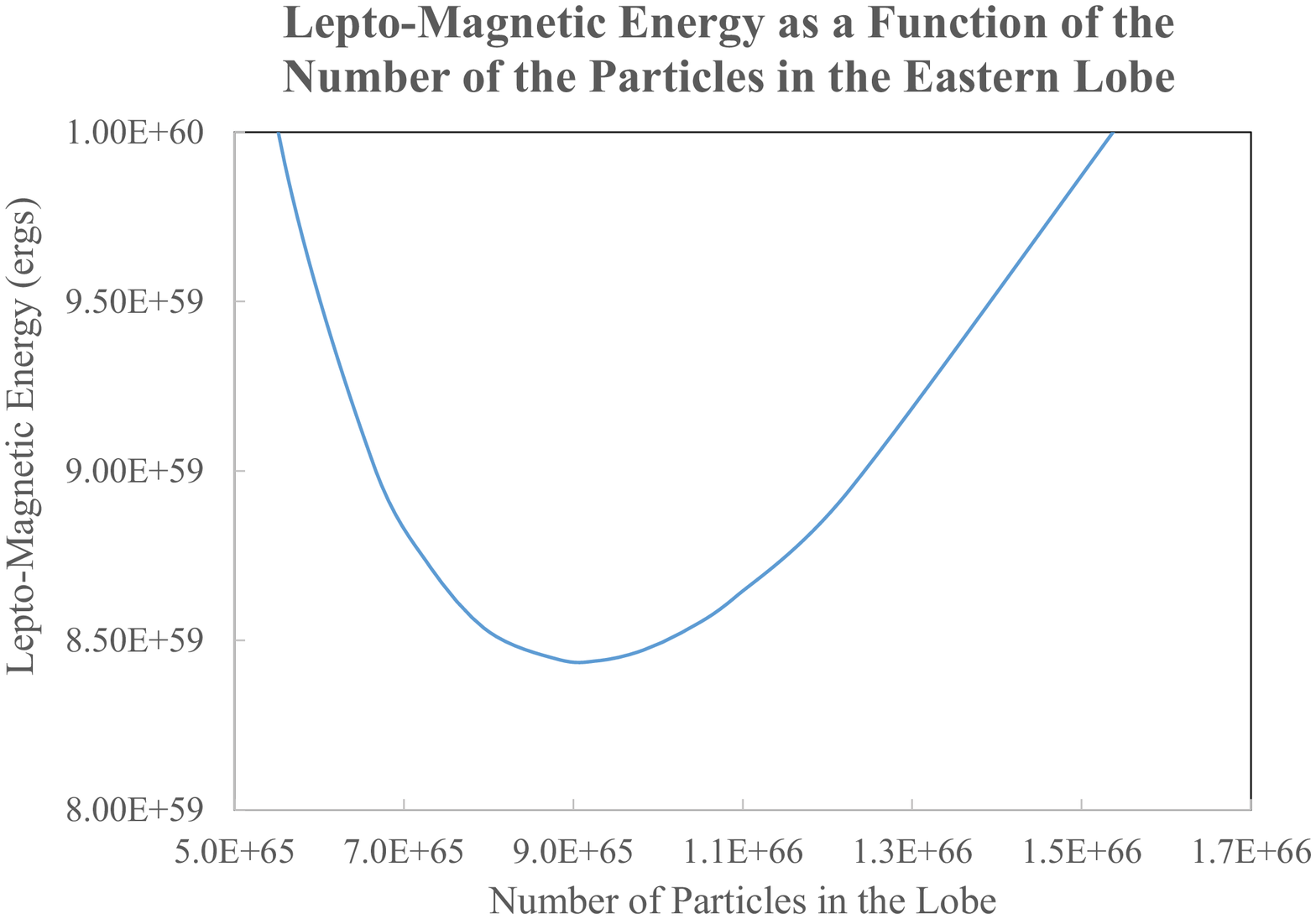}
\includegraphics[width= 0.45\textwidth,angle =0]{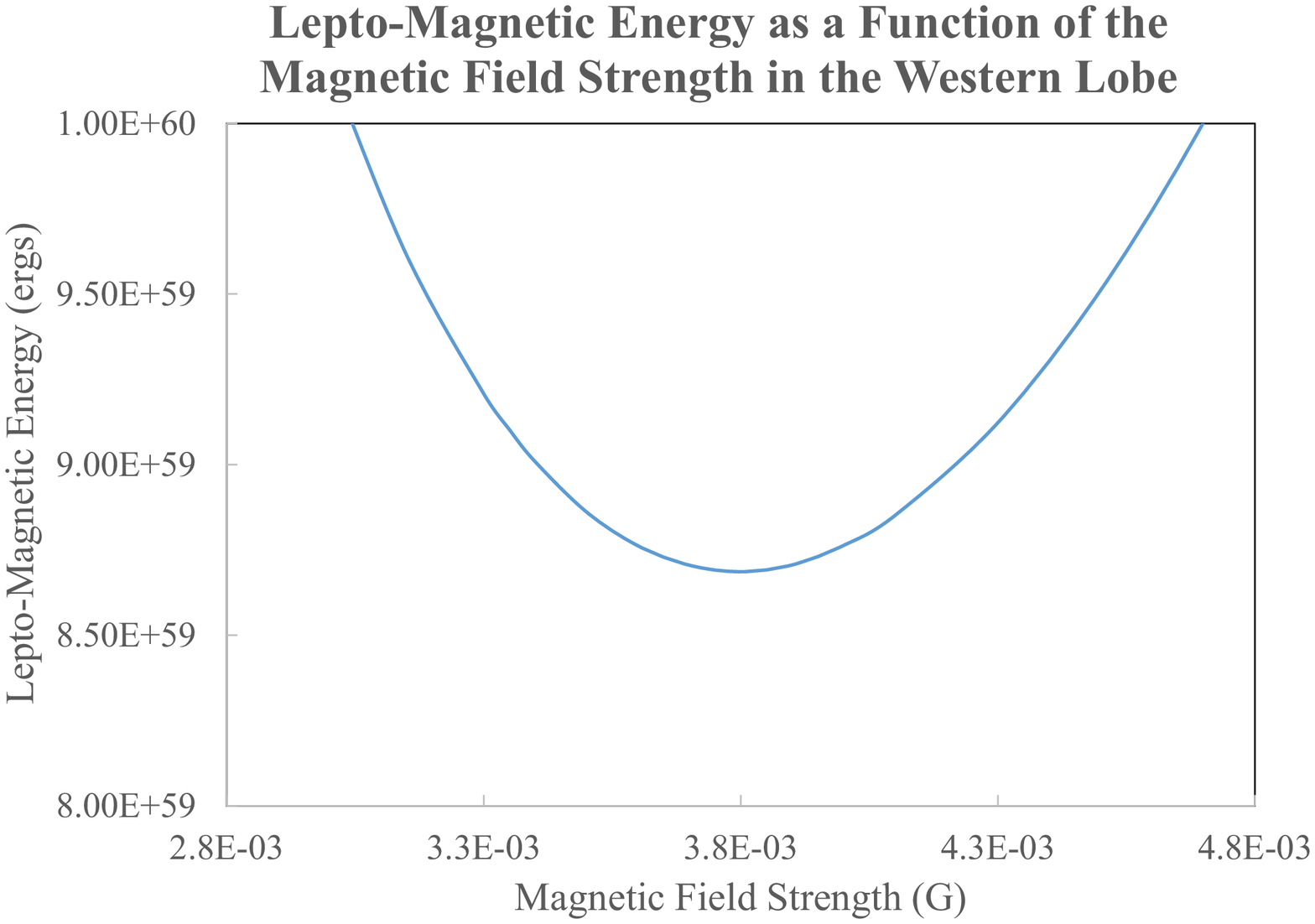}
\includegraphics[width= 0.45\textwidth,angle =0]{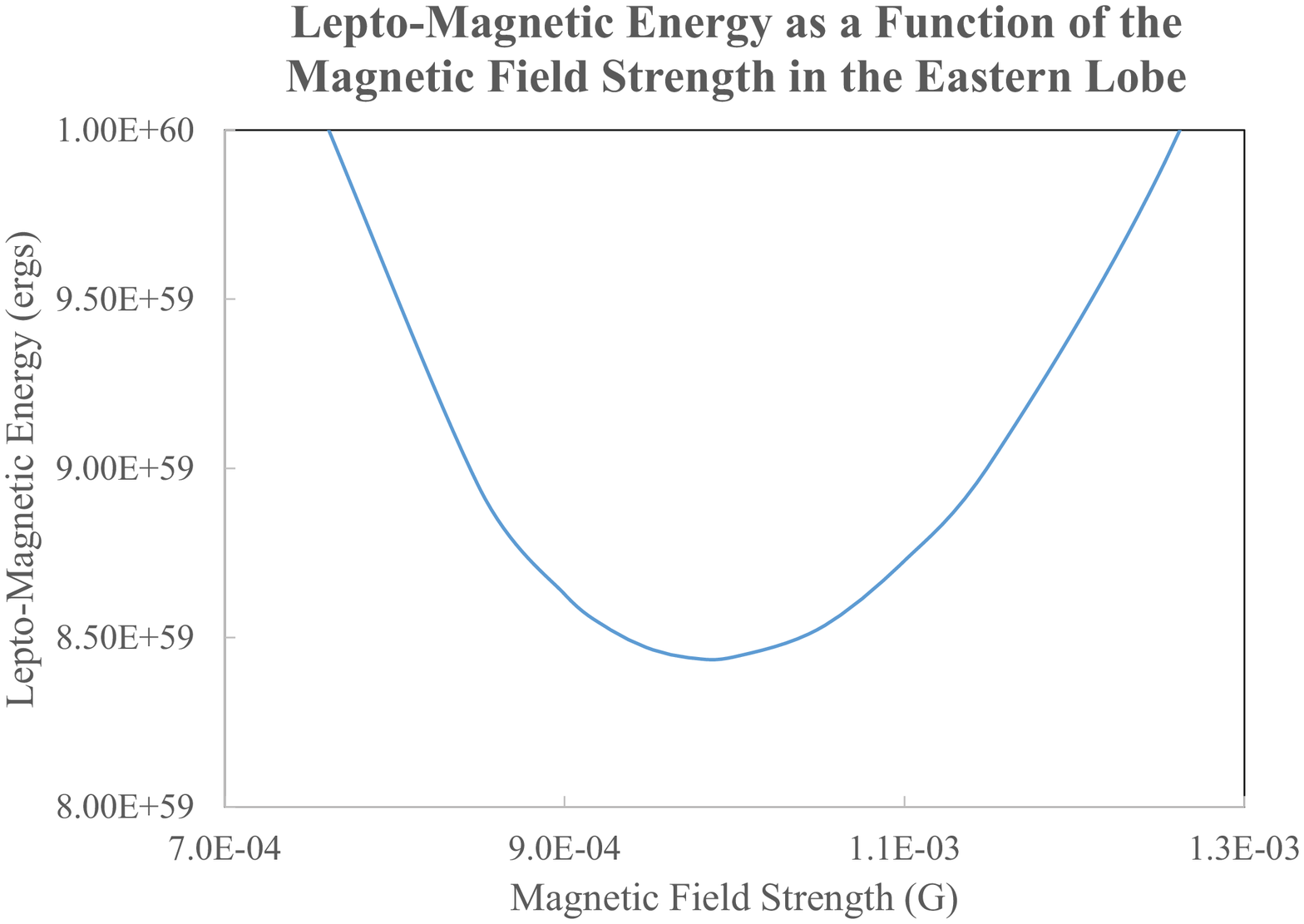}
\caption{The details of the solution with $E_{min}=m_{e}c^{2}$ for the fit to the radio spectrum in Figure 2. The curves in the panels are the graphical manifestation of the infinite 1-D set of solutions for each lobe. The top two panels are the dependence of $E(\mathrm{lm})$ on the plasmoid radius, $R$, for the western lobe (left) and eastern lobe, (right). The lower two rows are the dependence of $E(\mathrm{lm})$ on the total number of particles in the lobes and the magnetic field strength. The minimum  energy ($E(\mathrm{lm})$) in these solutions is the same in both lobes to within 3\%.}
\end{center}
\end{figure*}

\section{Physical Realizations of the Lobe SSA Power Laws}
The spherical homogeneous plasmoid models employed here have been described many times elsewhere as noted above. We compile the physical equations in Appendix A for ease of reference. More details can be found in \citet{pun20}, which will be referenced as needed in this abbreviated discussion. It has been argued that leptonic lobes are favored in the more luminous FRII (Fanaroff Riley II) radio sources \citep{cro18,fr74}. This model will be our starting point, but we will consider protonic lobes as well, subsequently. From Appendix A, mathematically, the theoretical determination of the flux density, $S_{\nu}$, of a spherical plasmoid depends on 7 parameters: $N_{\Gamma}$ (the rest frame normalization of the lepton number density spectrum), $B$ (rest frame magnetic field), $R$ (the radius of the sphere in the rest frame), $\alpha$ (the spectral index of the power law), $\delta$ (the Doppler factor), $E_{min}$ (the minimum energy of the lepton energy spectrum) and $E_{max}$ (the maximum energy of the lepton energy spectrum), yet there are only three constraints from the fit in Figure 2 for each lobe, $\overline{\tau}$, $\alpha$ and $S_{o}$, it is an under-determined system of equations. Most of the particles are at low energy, so the solutions are insensitive to $E_{max}$. For a slow moving lobe, $\delta=1$. $E_{min}$ is not constrained strongly by the data, so we start with an assumed $E_{min}=1$ then subsequently explore the ramifications of letting $E_{min}$ increase. One has four remaining unknowns, $ N_{\Gamma}$, $B$, $R$ and $\alpha$ in the physical models and three constraints from observations, $\overline{\tau}$, $\alpha$ and $S_{o}$, for each model. Thus, there are an infinite one-dimensional set of solutions in each lobe that results in the same spectral output.

\par The primary discriminant for a meaningful solution is the physical requirement:
 Over long periods of time, we assume that there is an approximate bilateral symmetry in terms of the energy ejected into each jet arm. Thus, we require the internal energy, $E(\mathrm{lm})$ (defined in Appendix A: Equation A(9)) to be approximately equal in both lobes in spite of the fact that the spectral index is not.

\par Figure 3 describes the physical parameters of the (infinite 1-D set of) models that produce the fit in Figure 2. The top panels show the dependence of $E(\mathrm{lm})$, the lepto-magnetic energy (internal lepton energy and magnetic energy), on $R$ (the lobe radius). The middle row displays the dependence of $E(\mathrm{lm})$ on the total number of particles in the lobe, $\mathcal{N}_{e}$. The bottom row shows the dependence on the turbulent magnetic field strength. The two independent lobe solutions have minimum energies ($E(\mathrm{lm})$) that agree within a few percent. Based on the lone physical requirement of the solution above (long term bilateral energy ejection symmetry), the minimum energy solution is significant. The minimum energy of the western (eastern) lobe is $E(\mathrm{lm})= 8.69 \times 10^{59}\, \rm{ergs}$ ($E(\mathrm{lm})= 8.44 \times 10^{59}\, \rm{ergs}$). We believe this agreement to be significant and not a coincidence of the mathematics for two reasons. First, this assertion is strongly favored based on the finding of \citet{pun12} that ejections from the X-ray binary, GRS~1915+105, evolve towards the minimum energy configuration as they expand away from the source. Secondly, and more on point, the detailed X-ray studies of FR~II radio sources indicate that the lobes are generally near minimum energy or perhaps slightly dominated by the particle energy \citep{ine17}.

\subsection{Converting Stored Lobe Energy into Jet Power}
There is a direct physical connection between $E(\mathrm{lm})$ and the long-term time-averaged power delivered to the radio lobes, $\overline{Q}$. If the time for the lobes to expand to their current separation is $T$, in the frame of reference of the quasar, then the intrinsic jet power is approximately
\begin{equation}
\overline{Q} \approx E(\mathrm{lm})/T
\end{equation}
This expression ignores the work done by the inflating radio source as it displaces the galactic medium. For powerful sources like 3C 298 and 3C 82, the lobes are over-pressurized relative to an elliptical galaxy environment and the work of expansion is negligible compared to $ E(\mathrm{lm})$ \citep{pun20,mat03}.

The mean lobe advance speed, $v_{\rm{adv}}$, can be found from the arm length asymmetry in Figure 1, if one assumes intrinsic bilateral symmetry and there is inconsequential disruption of jet propagation from interactions with the enveloping medium. The system appears very linear in Figure 1 and in the MERLIN observations at 1.66 GHz and 5 GHz \citep{spe89,lud98}. Thus, this might be a system where arm length ratio is a valid method of estimating $v_{\rm{adv}}$. The arm length ratio of approaching lobe, $L_{\rm{app}}$, to receding lobe, $L_{\rm{rec}}$ \citep{gin69,sch95}, is
\begin{eqnarray}
&& R = \frac{L_{\rm{app}}}{L_{\rm{rec}}} = \frac{1+(v_{\rm{adv}}/c) \cos{\theta}}{1-(v_{\rm{adv}}/c) \cos{\theta}} \;, \rm{where}\nonumber \\
&& L_{\rm{app}}=\frac{(v_{\rm{adv}}T)\sin{\theta}}{1-(v_{\rm{adv}}/c) \cos{\theta}} \;: \; L_{\rm{rec}}=\frac{(v_{\rm{adv}}T) \sin{\theta}}{1+(v_{\rm{adv}}/c)\cos{\theta}}\;,
\end{eqnarray}
where $T$ is the time measured in the quasar rest frame and projected length on the sky plane of an earth observer (corrected for cosmological effects) is $\mathcal{L}\equiv L_{\rm{app}}+L_{\rm{rec}}$. The arm length ratios were computed in \citet{sch95} using the lowest contour level in the images. We choose the second lowest contour in Figure 1, for the reasons given below, and find $R\approx 1.35$,
\begin{itemize}
\item the outer contour has an exaggerated distortion not seen in the more robust higher contour levels,
\item the second lowest contour is about one-half of the synthesized beam width (the resolution limit of the image) from the lowest contour and
\item this method applied to the MERLIN 1.66 GHz image in \citet{spe89} also yields $R\lesssim 1.4$.
\end{itemize}
Using the second lowest contour, the linear size of the source on the sky plane is 2\farcs8; using the lowest contour in Figure 1, the linear size is 3\farcs1. This angle is slightly larger than the 2\farcs5 found by \citet{spe89}; however, even with the best efforts to correct for ionospheric scattering, we expect some blurring in the image. We assume that the modest blurring is fairly uniform over the small image and does not significantly affect the $R$ estimate.

For a quasar, we can eliminate the possibility of a highly oblique line of sight (LOS) to the jet. The LOS to the jet in quasars is believed to be $<45^{\circ}$ with an average of $\approx 30^{\circ}$ \citep{bar89}. Since 3C 298 is very lobe dominated, one does not expect a LOS near the low end of this quasar range \citep{bro86}. In accord with this conclusion, a variability study consisting of thirty-eight observations at 408 MHz spread out over 19 years found no evidence of variability, which strongly disfavors a blazar-like LOS, $<10^{\circ}$ \citep{bon96}\footnote{Note that in Section 4, we remarked on possible modest variability above 8 GHz. Even though this is a tentative finding that is not verified, it does not indicate that a hidden blazar is necessarily present (as would be the case if a factor of a few variability were detected). In view of the thirty-eight observations at 408 MHz with the same telescope noted here, the modest variability at higher frequency requires rigorous verification to be considered more than tentative.}. In our adopted cosmology, at the distance of 3C298 the scale is 8.58 kpc per arc-second. From Equation (4), with $R$=1.35, and assuming $20^{\circ} < \rm{LOS} < 40^{\circ}$
\begin{equation}
v_{\rm{adv}} \approx 0.18 c \pm 0.02 c\;\rm{and}\; 9.59\times 10^{12} \ \rm{sec} < T < 2.22\times 10^{13} \ \rm{sec}\;.
\end{equation}
The value of $v_{\rm{adv}}$ in Equation (5) agrees with the mean value of $v_{\rm{adv}}$ estimated for a sample of relatively straight 3C CSS quasars \citep{pun20}. The large spread in $T$ in Equation (5) arises from the uncertainty in the LOS. Inserting Equation (5) into Equation (3) with $E(\mathrm{lm})=1.71\times 10^{60} \rm{ergs}$ from the minimum energy solutions in Figure 3 yields
\begin{equation}
7.73\times 10^{46}\rm{ergs/sec} < \overline{Q} < 1.79\times 10^{47} \ \rm{erg \ sec}^{-1}\;.
\end{equation}
\par We compare our estimate with the well-known estimates of $\overline{Q}$ that is based on the spectral luminosity at 151 MHz per steradian, $L_{151}$. This method is more suited to a relaxed classical double radio source and might not be applicable to CSS sources in general \citep{wil99}. $\overline{Q}$ is given as
a function of $f$ and $L_{151}$ in Figure 7 of \citet{wil99},
\begin{equation}
\overline{Q} \approx 3.8\times10^{45} f L_{151}^{6/7} \ \rm{erg \ s}^{-1}\;,
\end{equation}
The parameter $f$ was introduced to account for deviations from 100\% filling factor, minimum energy, a low frequency cut off at 10
MHz, the jet axis at $60^{\circ}$ to the line of sight, and no
protonic contribution as well as energy lost expanding the lobe into the external medium,
back flow from the head of the lobe and kinetic turbulence \citep{wil99}.
The exponent on $f$ in Equation (7) is 1, not 3/2 as was previously reported \citep{pun18}. The quantity
$f$ has been estimated to lie in the range of $10<f<20$ for most FRII radio sources
\citep{blu00}. Using $L_{151}$ from Table 1 and $10<f<20$
\begin{equation}
\overline{Q} = 2.18\pm 0.71\times 10^{47} \ \rm{erg \ s}^{-1} \;.
\end{equation}
The uncertainty in Equation (8) corresponds to the range of $10<f<20$. The marginal agreement between Equations (6) and (8) is reasonable considering all the unknown dynamics. De-projecting a linear size of 2\farcs5 with $20^{\circ}<\rm{LOS}<40^{\circ}$ yields $\sim 30-60$ kpc. The larger value of Equation (8) compared to Equation (6) does support the notion that compact sources will tend to be over-luminous relative to their jet power due to interactions with the galactic environment \citep{bar96,wil99}. Formally, there is an overlap in the values of $\overline{Q}$ if the uncertainty is considered which is remarkable considering that the two methods are independent and are based on different assumptions.
\par The result is not strongly dependent on the choice of the second lowest contour of Figure 1 for the computation of $R$ as opposed to the lowest contour in \citep{sch95}. Using the lowest contour in Figure 1, $R=1.38$ and Equation (5) becomes
\begin{equation}
v_{\rm{adv}} \approx 0.19 c \pm 0.02 c \ \ \ \ \rm{and} \ \ \ \  1.02\times 10^{13} \ \rm{s} \ < T < \ 2.35\times 10^{13} \ \rm{s}\;.
\end{equation}

\subsection{Assessing the $E_{min}\approx m_{e}c^{2}$ Assumption and Protonic Content}
We started the analysis of the solution space with two basic assumptions:
\begin{itemize}
\item $E_{min}(\rm{West\, Lobe}) = E_{min}(\rm{East\, Lobe}) = m_{e}c^{2}$ and
\item The plasma is predominantly positronic, not protonic.
\end{itemize}
In this section, we examine the solutions when these assumptions are violated.

\par We explored the possibility of altering the low energy cutoff, by arbitrarily setting $E_{min}(\rm{West\, Lobe}) = E_{min}(\rm{East\, Lobe}) = 5m_{e}c^{2}$.
In this case $E_{\rm{lm}}(\rm{East\, Lobe}) =1.45 E_{\rm{lm}}(\rm{West\, Lobe})$ in their minimum energy states. This ratio continues to increase as one raises  $E_{min}$. For a system that requires long term bilateral energy flux symmetry, raising $E_{min}$ arbitrarily at least the west lobe must start deviating significantly from its minimum energy state. The solution that we have explored in Figure 3 has $E_{min}(\rm{West\, Lobe}) = E_{min}(\rm{East\, Lobe}) = m_{e}c^{2}$. The solution has bilateral symmetry in energy content, $E_{\rm{lm}}(\rm{West\, Lobe}) \approx E_{\rm{lm}}(\rm{East\, Lobe})$ if both lobes are near their minimum energy state in Figure 3. This is a natural state of a relaxed plasmoid and radio lobes in particular \citep{pun12,ine17}. There are no unexplained spectral breaks or low energy cutoffs in the lepton spectra. Thus, we consider this the most plausible solution, but we cannot rule out other configurations.

\par If we assume protonic matter instead of positronic matter, the spectral fit in Figure 2 remains unchanged and is created by electrons in the magnetic field. A significant thermal proton population in FR II radio lobes has been argued to be implausible based on pressure balance with the enveloping medium \citep{cro18}. These are therefore cold protons. In this case, consider the mass stored in the lobes based on Figure 3 for the minimum energy, minimum $E(\mathrm{lm})$, solution, $M_{\rm{lobe}} = 1.38\times 10^{8} M_{\odot}$. For the average $T$ in Equation (5), this mass corresponds to 274 $M_{\odot}$ on average that must be ejected a year in order to fill the lobes with protonic matter. For the black hole mass $M_{bh} \approx 3\times 10^{9}M_{\odot}$ from \citet{vay17}, the lobes would need to be fed at $>4$ times the Eddington rate in order to be populated with protonic matter. In Section 9, we find a viral $M_{bh}$ even smaller than this value. As with 3C 82, the baryon ejection rate must greatly exceed the accretion rate for protonic lobes to exist. We come to the conclusion that the protonic lobe solutions are not highly plausible.

\begin{figure}
\begin{center}
\includegraphics[width= 0.85\textwidth]{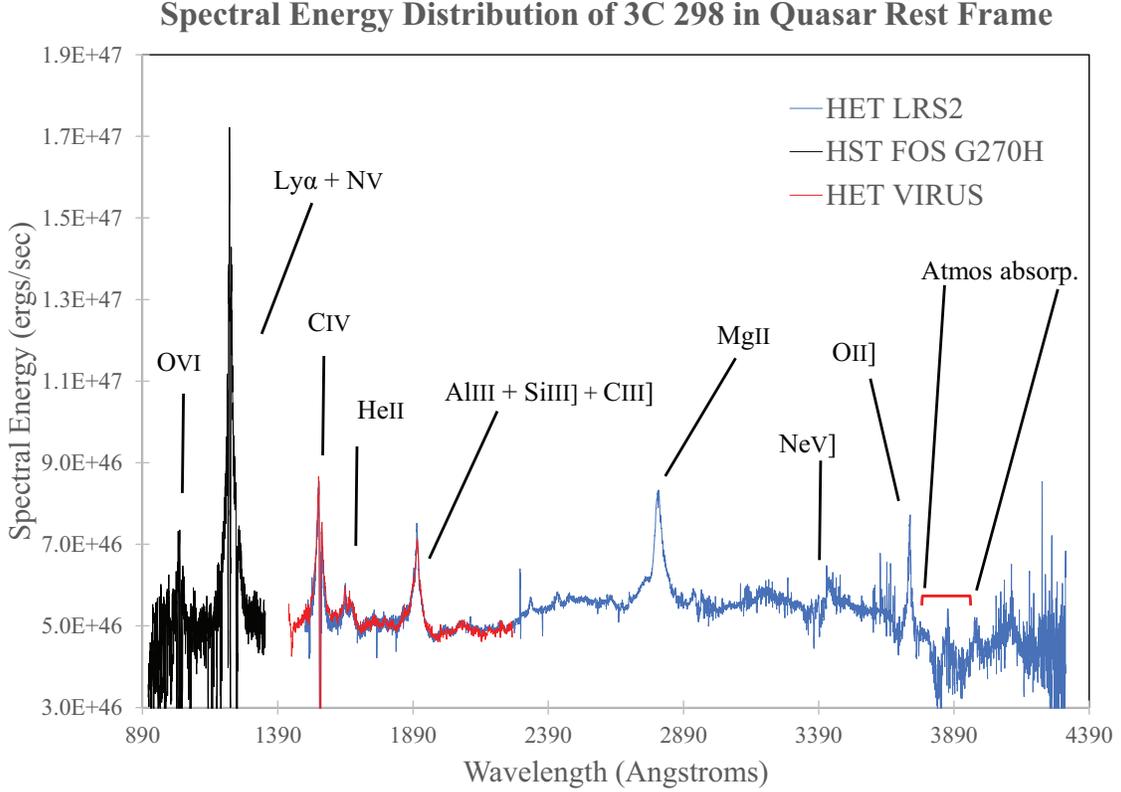}

\caption{The SED of 3C 298 showing the prominent emission lines. The data are presented in terms of frequency in the cosmological frame of reference of the quasar. The spectral energy is computed in the quasar reference frame in terms of the spectral luminosity, $L_{\nu}$, as $\nu L_{\nu}$. The SED was corrected for Galactic extinction. The SED combines three spectra, the HST archival spectrum, and the new HET LRS2 and VIRUS spectra obtained for the purposes of this paper. The broad emission lines are presented in detail with fits in Appendix B.}
\end{center}
\end{figure}

\section{The Ultraviolet Spectrum}
Our motivating interest in the UV spectrum is to characterize the possibility of a strong blue wing in CIV that might be indicative of a high ionization wind. An exaggerated blue wing in CIV is evident in the high signal-to-noise ratio spectrum of \citet{and87}. Unfortunately, these data are not flux calibrated. Two spectrographs were employed on the upgraded 10~m Hobby-Eberly Telescope (HET, \citealt{hill21}) in order to capture a spectrum from the blue wing of CIV (at 3500~\AA) to [OII] (at 9100~\AA). The short wavelength portion of the spectrum is covered by Visible Integral-Field Replicable Unit Spectrograph (VIRUS, \citealt{hill21}).
The long wavelength portion of the spectrum is covered by an observation with the new Low Resolution Spectrograph 2 (LRS2, \citealt{cho16}, Hill et al. 2022 in prep.) to obtain spectra on April 28,
UT. We used both units of the integral field spectrograph, LRS2-B and LRS2-R with the target switched between them on consecutive exposures. Each unit is fed by an integral field unit with 6x12 square arcsec field of view, 0.6$''$ spatial elements, and full fill-factor, and has two spectral channels. LRS2-B has two channels, UV (3700~\AA ~- 4700~\AA) and Orange (4600~\AA ~- 7000~\AA), observed simultaneously. LRS2-R also has two channels, Red (6500~\AA ~- 8470~\AA) and Far-red (8230~\AA ~- 10500~\AA).
The observations were split into two 1200 second exposures, one for LRS2-B and one for LRS2-R. The image size was 1\farcs38 full-width half-maximum (FWHM).
The spectra from each of the four channels were processed independently using the HET LRS2 pipeline, Panacea\footnote{https://github.com/grzeimann/Panacea} (Zeimann et al. 2022, in prep.). The LRS2 data processing followed that for 3C82 and details can be found in \citet{pun20}.

VIRUS is a highly replicated, fiber-fed integral field spectrograph. The entire VIRUS instrument consists of 156 integral field low-resolution (resolving power, $R\sim$800) spectrographs, arrayed in 78 pairs. Each of the 78 VIRUS units is fed by an integral field unit (IFU) with 448 $1\farcs5$ fibers covering 51$'' \times$51$''$ area.
The spectrographs have a fixed wavelength coverage of 3500 $< \lambda <$ 5500~\AA.
The IFUs are arrayed in a grid pattern in the 18$'$ field of view of HET with a $\sim$1$/$4.5 fill-factor and VIRUS captures 34,944 spectra for each exposure at full compliment. An observation consists of three dithered exposures with small position offsets to fill in the gaps between the fibers in the IFUs.
3C298 was observed on one of the VIRUS IFUs on June 21, 2019 UT for 3x360 seconds total exposure time. Image quality was 1.55$\arcsec$ FWHM.
The VIRUS data were reduced with the pipeline Remedy\footnote{https://github.com/grzeimann/Remedy} (Zeiman et al. 2022, in prep.). There are multiple field stars in the VIRUS field of view with {\it g}-band calibrations from Pan-STARRS \citep{cha19} Data Release 2 Stacked Object Catalog
(PS-1 DR2, \citealt{fle20}) that provide direct flux calibration of the VIRUS data; since there is overlap between the spectra from the two spectrographs, the VIRUS observation anchors the flux density calibration for the long wavelength portion of the spectrum from LRS2. With a 10\% adjustment in the flux density of the LRS2 spectra, the independently calibrated spectra from VIRUS and LRS2-B agree to 1\% in the overlap region (3700 $< \lambda <$ 5500~\AA).
\par The spectrum is presented in the form of a spectral energy distribution in Figure 4, which is corrected for Galactic extinction. The extinction values in the NASA Extragalactic Database (NED) were used in a \citet{car89} model; $A_{V}=0.08$ and $R_{V}=3.1$. Figure 4 also includes the Hubble Space Telescope (HST) Faint Object Spectrograph, G270H spectral data (from August 15, 1996) downloaded from NED \citep{bec02}.

\begin{table}
\caption{Ultraviolet Broad Emission Line Fits}
\tiny{\begin{tabular}{ccccccccccc} \tableline\rule{0mm}{3mm}
  Line & VBC & VBC & VBC & BLUE & BLUE & BLUE & BC & BC & Total BEL \\
   &  Peak\tablenotemark{\tiny{a}} & FWHM & Luminosity & Peak\tablenotemark{\tiny{a}} & FWHM &Luminosity & FWHM &Luminosity & Luminosity  \\
    &  km~s$^{-1}$ &  km~s$^{-1}$ & ergs~s$^{-1}$ & km~s$^{-1}$ &km~s$^{-1}$ & ergs~s$^{-1}$ & km~s$^{-1}$  & ergs~s$^{-1}$ & ergs~s$^{-1}$  \\
\tableline \rule{0mm}{3mm}
 Ly$\alpha$ $\lambda1216$  & 5246 & 9897 &  $6.25 \times 10^{44}$ & -3381 & 8208  & $8.18 \times 10^{44}$ & 4380 & $9.57 \times 10^{44}$ & $2.40 \times 10^{45}$ \\
 NV$\lambda1240$ & 5212 & 9897 &  $7.70 \times 10^{43}$ & -5075  & 3425  & $1.11 \times 10^{44}$ & 4800 & $1.09 \times 10^{44}$ & $2.97 \times 10^{44}$ \\
 CIV$\lambda1549$ & 2130 & 15000 &  $3.25 \times 10^{44}$& -2460  & 5907 & $2.75 \times 10^{44}$ & 3890 & $3.19 \times 10^{44}$ & $9.20 \times 10^{44}$ \\
 HeII$\lambda1640$ & 2202 & 15000 &  $1.77 \times 10^{44}$& -1573 & 6376 & $2.63 \times 10^{43}$ & 3890 & $3.67\times 10^{43}$ & $2.12\times 10^{44}$ \\
 AlIII$\lambda1854$ & ... & ... & ...  & ... & ... & ...  & 2810 & $1.13 \times 10^{43}$ & $1.13 \times 10^{43}$ \\
 AlIII$\lambda1863$ & ... & ... & ... & ... & ... & ...  & 2810  & $8.97 \times 10^{42}$ & $8.97 \times 10^{42}$ \\
  SiIII]$\lambda1892$ & ...& ... & ... & ... & ... & ...  & 4014  & $7.88 \times 10^{43}$ & $7.88 \times 10^{43}$ \\
 CIII]$\lambda1909$ & -157 & 6000 & $1.08 \times 10^{44}$ & ... & ... & ...  & 4000  & $7.95 \times 10^{43}$ & $1.87 \times 10^{44}$ \\
 MgII$\lambda2799$ & 182 & 10556 & $2.83 \times 10^{44}$ & ... & ... & ...  & 2795  & $1.78 \times 10^{44}$ & $4.60 \times 10^{44}$
\end{tabular}}
\tablenotetext{a}{Peak of the fitted Gaussian component in km~sec$^{-1}$ relative to the quasar rest frame. A positive value is a redshift.}
\end{table}
Table 2 provides standard three Gaussian component decompositions of the broad emission lines (BELs): a broad component, ``BC", \citep[also called the intermediate broad line or IBL;][]{bro94}, an often redshifted, ``VBC" \citep[very broad component following][]{sul00} and ``BLUE" a blueshifted excess \citep{bro96,mar96,sul00}. The table is organized as follows. The line designation is defined in the first column. The next three columns define the properties of the Gaussian fit to the VBC, the shift of the Gaussian peak relative to the vacuum wavelength in km~sec$^{-1}$, followed by the FWHM and line luminosity. Columns (5)-(7) are the same for the BLUE. The BC FWHM and luminosity are columns (8) and (9). The last column is the total luminosity of the BEL.

The decompositions are shown in Appendix B, after continuum subtraction. The line fitting is challenged by various issues.
\begin{itemize}
\item The Ly$\alpha$, NV blend has very strong narrow absorption features throughout. As in \citet{bec02}, we incorporated significant spectral smoothing to accentuate the line profiles. The strong absorption near zero velocity makes segregating the narrow line component impossible. Thus, the line decomposition is not exact, but we do not require precise knowledge of the flux associated with each component.
\item The CIV, HeII blend also has strong zero velocity absorption. Again, no narrow CIV line can be decomposed from the complex. However, there are mitigating factors. The absorption is relatively narrow (FWHM $\sim$ 1000 km/s) so that only the central part of the profile is affected. Most of the CIV core and of the wings are free from contaminants, so that the broad component profile appears well-defined. To take into account the deep absorption and the blending of HeII, we resort to the above-mentioned, multi-component fitting that has proven to be appropriate in case of prominent CIV without strong absorption. This permits us to extrapolate the observed profile across the absorption seen in 3C 298. The line flux uncertainty is also well-constrained, since the narrow component of CIV is weak, accounting for 2-20\% of the total CIV flux in quasars \citep{mar96,sul99}.

\item The MgII background is highly complex and determining the continuum level is not trivial. There is a rather typical, strong FeII$\lambda$2235-2670 feature and a strong Balmer continuum. The high spectral resolution, high signal-to-noise ratio, LRS2 spectrum was used for fitting these features that mask the continuum level.
\end{itemize}
In spite of these limitations to the fitting process, a few general conditions are evident. There is a significant blue excess in the Ly$\alpha$, NV, CIV, and HeII BELs; no such excess appears in SiIII], CIII], and MgII. In radio loud quasars, even though BLUE of CIV is often detectable, it is usually significantly weaker than the red VBC \citep{pun10}. Yet, the BLUE of the CIV broad line in 3C298 is nearly as luminous as the red VBC, which is extreme for a radio loud quasar.
\begin{table}
\caption{Emission Line Strengths}
\tiny{\begin{tabular}{cccccccc} \tableline\rule{0mm}{3mm}
  Line  & Total      & Total                     & HST Composite\tablenotemark{\tiny{a}}           & EW       & HST Composite\tablenotemark{\tiny{a}} & EW \\
        & Luminosity & Luminosity                &Luminosity               &          & EW              &  Ratio \\
        & ergs~s$^{-1}$ & Relative to Ly$\alpha$ & Relative to Ly$\alpha$  & \AA    & \AA           &  \\
\tableline \rule{0mm}{3mm}
 Ly$\alpha$ $\lambda1216$  & $2.40 \times 10^{45}$ & 1  & 1 & 59.3 & 91.8 & 0.65 \\
 NV$\lambda1240$ & $2.97 \times 10^{44}$  & 0.12  & 0.20 & 7.3 & 18.5 & 0.39 \\
 CIV$\lambda1549$ & $9.20 \times 10^{44}$  & 0.38 & 0.48 & 29.0 & 58.0 & 0.50 \\
 HeII$\lambda1640$ &$2.57\times 10^{44}$  & 0.11 & 0.01 &7.9 & 1.5\tablenotemark{\tiny{e}} & 5.7\tablenotemark{\tiny{e}} \\
 AlIII &$2.03 \times 10^{43}$  & 0.009 & 0.015 & 0.8 & 2.2 & 0.36 \\
 SiIII]+CIII] &$3.00 \times 10^{44}$ & 0.13 & 0.13  & 12.1  & 19.7 & 0.62\\
FeII$\lambda$2235-2670 &$8.58 \times 10^{44}$& 0.36 & \tablenotemark{\tiny{b}}  & $\sim 43$  & \tablenotemark{\tiny{b}} & \tablenotemark{\tiny{b}}\\
 MgII$\lambda2799$ &$4.86 \times 10^{44}$& 0.20 & 0.23  & 25.6  & 51.7 & 0.49\\
 H$\alpha$+NII &$1.86 \times 10^{45}$ \tablenotemark{\tiny{c}} & 0.78 & 0.77\tablenotemark{\tiny{d}}  & ...  & ... &\\
 \hline
 Sum &$7.40 \times 10^{45}$  & 3.10 & N/A  & N/A  & N/A & N/A

\end{tabular}}
\tablenotetext{a}{\citet{tel02}}
\tablenotetext{b}{Not computed in \citet{tel02}}
\tablenotetext{c}{\citet{hir03}}
\tablenotetext{d}{Not an HST composite result. Value used in \citet{cel97} for reference purposes.}
\tablenotetext{e}{\citet{tel02} did not deblend the red wing of CIV and the HeII profile, attributing a large fraction of the flux between the two lines to a $\lambda$ 1600 \AA\ feature. The resulting HeII equivalent width is therefore most likely severely underestimated.}
\end{table}
\par Table 3 lists the emission line strengths. We compare these results to the line strengths from an HST composite based on 184 quasars \citep{tel02}. In the second column, the line luminosity is slightly larger for some lines (SiIII]+CIII], MgII and HeII) than in Table 2 because narrow line contributions were added (since the \citet{tel02} fits were total line luminosity and we want to compare similar quantities) . Also some lines are blended \citep{tel02}. These changes are for comparison purposes to Table 2 of \citep{tel02}. A comparison of the line luminosity ratios in columns (3) and (4) reveals that the line strength ratios are similar to the HST composite values, which have estimated uncertainties in Table 2 of \citet{tel02} of less than~$10\%$. The last three columns compare the rest frame equivalent widths, EW, of 3C 298 and the HST composite, no uncertainties were given in \citet{tel02}. The last column indicates that most of the EWs are 40\%-60\% of the HST composite EWs. The broad line region is much smaller than can be resolved with existing telescopes, so this is an intrinsic difference between 3C 298 and most other quasars. The physical origin of the weak BELs are discussed in Section 8. FeII$\lambda$2235-2670 is quite prominent in the SED of Figure 3. From Table 3, the line strength ratio L(FeII$\lambda$2235-2670)/L(MgII)$\approx 1.8$, close to the value of 2.2 determined from a Large bright Quasar Survey Composite \citep{bal04}.

\par Table 3 indicates that the line ratios of 3C 298 are typical of most quasars, but the line strengths are weak compared to the strength of the UV continuum (the big blue bump) \citep{tel02,zhe97}. This result is consistent with a disproportionately weak ionizing continuum relative to the UV continuum. We explore this issue in Section 8.
\par Finally, in spite of the small EWs for most BELs, HeII is quite strong compared to a typical quasar. A similar powerful CSS quasar, 3C82, also has over-luminous HeII emission \citep{pun20}.
\section{The Bolometric Luminosity of the Accretion Flow}
We have sufficient information to construct an accurate SED for the accretion flow, and we have numerous broad line estimates, so we directly compute the bolometric luminosity of the accretion flow, $L_{\rm{bol}}$. We do not include reprocessed radiation in the infrared from molecular clouds located far from the active nucleus since doing so would be double counting the thermal accretion emission that is reprocessed at mid-latitudes \citep{dav11}. There are three major contributors to $L_{\rm{bol}}$:
\begin{enumerate}
\item The thermal emission from the accretion flow which has a broad peak in the optical and UV
\item The broad emission lines
\item The X-ray power law (corona) produced by the accretion flow.
\end{enumerate}

\begin{figure}
\begin{center}
\includegraphics[width= 0.85\textwidth]{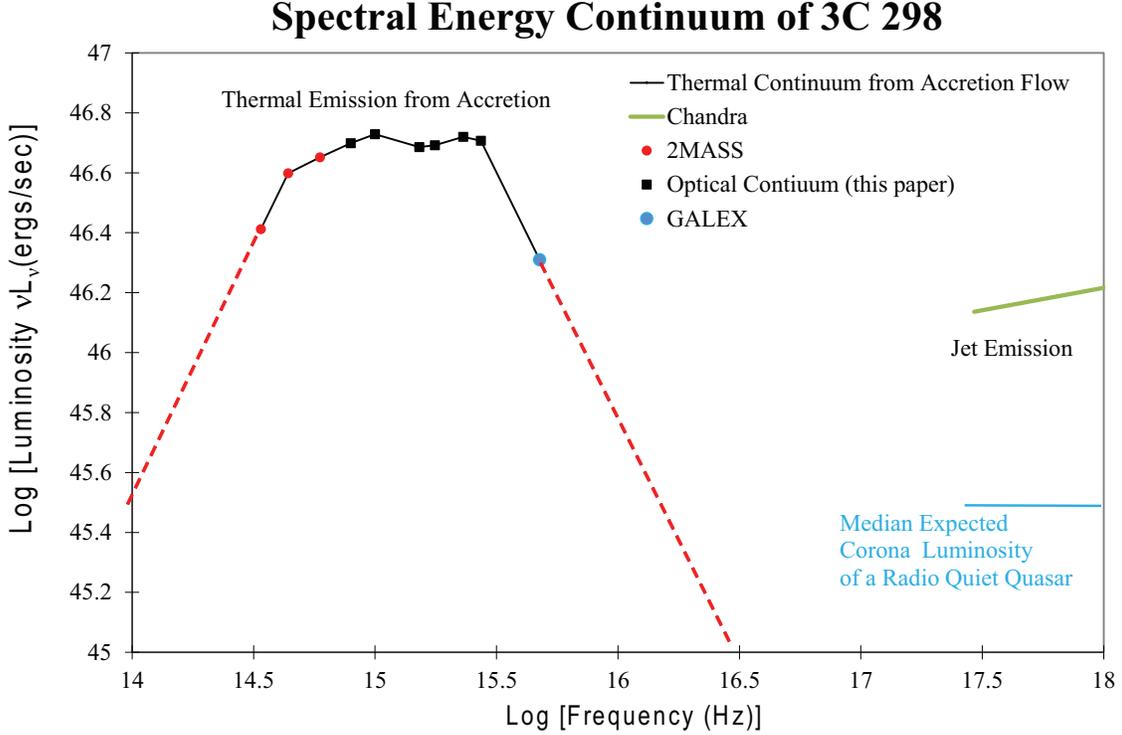}

\caption{The continuum SED of 3C 298 obtained from the SED in Figure 4, after the emission lines (fits given in Tables 2 and 3) were removed. We also include archival 2MASS and GALEX data. The SED contains a well-defined thermal component with steep declines in the IR and EUV. These steep declines are extrapolated by the dashed red lines. The \emph{Chandra} luminosity is much higher than that expected from an accretion disk corona. This favors a powerful jet as the primary source of the emission.}
\end{center}
\end{figure}
\subsection{The Thermal Continuum}Figure 5 includes data from the continuum of the SED in Figure 4. To construct the BEL fits in Table 2, we determined the continuum levels at 1300 \AA, 1700\AA, 1970 \AA, and 3000~\AA. These SED points, plus two others estimated from Figure 4 at 1100 \AA\, and 3700 \AA\, appear in the continuum SED in Figure 5. These six points accurately define the peak of the SED. To estimate the red end of the thermal SED, we added three photometric measurements from the Two Micron Survey (2MASS) that clearly delineate the 1 micron dip that is characteristic of quasar spectra \citep{skr06}. The J and H band points are a perfect extrapolation of the power law from the UV found from our HET observations. We extrapolate the power law plunge into the 1 micron dip with a dashed red line.
\par There are two Galaxy Evolutionary Explorer (GALEX) far UV (FUV) photometry points at 1539 \AA\ (631~\AA, in the quasar rest frame) in the GALEX GR6/7 Data Release\footnote{\url{https://galex.stsci.edu/GR6}}. The All Sky Imaging Survey reported a far UV AB magnitude of $m_{\rm{AB}}(FUV) =19.39 \pm 0.12$ and the Medium Imaging Survey found $m_{\rm{AB}}(FUV) =19.17 \pm 0.04$ \citep{mor07}. There are no flags on these data, the object is detected at high signal-to-noise ratio in both images (as reflected in the uncertainties in the magnitudes), there is no source confusion in the images and the source is not near the edge of the field of view.
These $m_{\rm{AB}}(FUV)$ values are converted to flux densities with the formula from the GALEX Guest Investigator Web Site \footnote{\url{https://asd.gsfc.nasa.gov/archive/galex}}
\begin{equation}
m_{\rm{AB}}(FUV) = -2.5 \log_{10}\left[\frac{F_{\lambda}(1539 \ {\rm \AA})}{1.40 \times 10^{-15} \ \rm{erg \ s}^{-1} \ \rm{cm}^{-2} \ \AA^{-1}}\right] + 18.82 \;.
\end{equation}
We take the average of the two flux densities from Equation (10), corresponding to the two GALEX measurements to find (after correcting for Galactic extinction),
\begin{equation}
F_{\lambda}(1539 \ {\rm \AA})=1.12 \pm 0.11 \times 10^{-15} \ \rm{erg \ s}^{-1} \ \rm{cm}^{-2} \ \AA^{-1} \;.
\end{equation}
We corrected this value for Ly$\alpha$ forest absorption and the broad emission lines in the wide photometric window. The FWHM of the FUV band is 228 \AA\, according to the GALEX Guest Investigator Web Site. In the quasar rest frame the half-maxima points are at $631 \pm 47$ \AA. In a study of HST quasar spectra, \citet{ste14} reported the presence of strong OIV and OV BELs in this region. In their composite spectra, these BELs comprise about 12\% of the flux in this band. This effect is much larger than absorption by the Ly$\alpha$ forest, since an observed wavelength of 1539 \AA\, corresponds to a Ly$\alpha$ photon at a redshift of only, $z \sim 0.25$. The absorption is estimated to typically be $< 2\%$ at $z=0.25$ \citep{zhe97}. Combining the two effects, we estimate that the GALEX FUV photometry will overestimate the flux in the pass-band by $\approx 10\%$. After applying this correction to Equation (11), we estimate a luminosity (in terms of the spectral luminosity $L_{\lambda}(\lambda)$),
\begin{equation}
\lambda L_{\lambda}(631 \ {\rm \AA})= 2.05\pm 0.21\times 10^{46} \ \rm{erg \ s}^{-1} \;.
\end{equation}
This point was added to the SED in Figure 5. The extreme ultraviolet (EUV) spectral index from 1100~\AA, (HST) to 631~\AA\, (GALEX), $\alpha_{\rm{EUV}}$, is defined in frequency space, $F_{\nu} \propto \nu^{-\alpha_{\rm{EUV}}}$ (from frequency ($2.73 \times 10^{15}$ Hz to $4.75 \times 10^{15}$ Hz). In Figure 5,
\begin{equation}
\alpha_{\rm{EUV}} = 2.64 \pm 0.18 \;.
\end{equation}
The red dashed line in Figure 5 extrapolates this power law to higher frequency.

\subsection{The X-ray Data}
The SED in Figure 5 includes the \emph{Chandra} X-ray data \citep{sie08}. The X-ray power law can arise from either a jet or the corona of an accretion disk in a powerful radio loud quasar. For perspective, if the jet were weak (a radio quiet quasar), based on the 1100~\AA\, edge of the big blue bump, one expects a corona level approximately where the light blue line is plotted in the lower right \citep{sha11,lao97}, i.e., the \emph{Chandra} flux levels are far in excess of what is expected of a corona. The X-ray luminosity in 3C 298 is large even compared to 3CR quasars with the largest 178 MHz luminosity \citet{sal08}. The total \emph{Chandra} luminosity from 0.5 keV to 10 keV (1.2 keV to 24 keV in the cosmological rest frame of the quasar) is $5.17\times 10^{46}$ erg s$^{-1}$ \citep{sie08}. This high-energy component is not only clearly X-ray emission of jetted origin, but is extremely large for a quasar, even most blazars; it is comparable to that of the brightest gamma-ray blazars \citep{ghi10}. Based on the light blue line in Figure 5, we conclude that the corona emission is masked by a much brighter jet. Due to this circumstance we need to estimate this undetected portion of the SED using composite information from other quasars. We estimate that $10\%$ of the big blue bump continuum luminosity is a reasonable composite X-ray luminosity of the accretion flow corona \citep{sha11,lao97}.
\par The jet X-ray luminosity is too large to be created by shocks from a frustrated jet in the CSS quasar \citep{ode17}. The X-ray emission is stronger in 3CR quasars than 3CR radio galaxies of similar radio power; this behavior has been attributed to modest Doppler beaming \citep{sal08}. This model appears plausible for 3C 298 based on the asymmetry of the inner jet in the VLBI images at 1.66 GHz and 5 GHz \citep{fan02}.
\subsection{Calculation of $L_{\mathrm{bol}}$}
As discussed in the introduction to this section $L_{\rm{bol}}$ has three components. The first, $L(\rm{thermal\; from \; accretion})$ is computed from direct integration of the broken power-law approximation to the continuum in the optical, UV and EUV in Figure 5. The second component is the broad emission luminosity. The broad line luminosity was estimated for quasars by $L(\rm{BELs}) \approx 5.6 L(\rm{Ly}\alpha)$ in \citep{cel97}. From the value of $L(\rm{Ly}\alpha)$ in Table 3,
$L(\rm{BELs}) \approx 1.34 \times 10^{46}\rm{ergs/sec}$. This seems reasonable based on the sum of the BEL luminosities in the last row of Table 3. The third component is the energy radiated by high energy electrons in a corona above the disk. In radio loud quasars this is not necessarily detected because it is likely masked by the X-ray emission from the radio jet. Thus, we invoked a crude approximation of 10\% of  $L(\rm{thermal\; from \; accretion})$ from the observations of radio quiet quasars (we assume similar coronae to first order) where it is detected.
Thus, we have
\begin{eqnarray}
&& L_{\rm{bol}} = L(\rm{thermal\; from \; accretion})  + L(\rm{BELs})+ L(\rm{corona}) \nonumber\\
&& \approx 1.29 \times 10^{47}\rm{ergs/sec} +1.34\times 10^{46}\rm{ergs/sec}+1.29 \times 10^{46}\rm{ergs/sec}= 1.55 \times 10^{47}\rm{ergs/sec} \;.
\end{eqnarray}
\par One can try to model the accretion disk SED with a mixture of thin disks and winds. For example, \citet{lao14} was able to find an explanation of the SED turnover at $1100 \AA$ in quasars. There are numerical attempts to model quasar disks with slim disk models \citep{sad11}. All of these models can be used to estimate accretion rates and central black hole masses. However, none of them can explain the steep EUV continuum in powerful radio loud quasars \citep{pun15}. Thus, we are motivated to explore this in Section 8.

\section{The EUV Deficit and the Weak Emission Lines}
In this section, we synthesize three findings of this paper:
\begin{itemize}
\item the enormous $\overline{Q}$ in Equation (6),
\item the steep EUV continuum in Equation (13) and Figure 5,
\item and the small BEL EWs in Table 3.
\end{itemize}
\begin{figure}
\begin{center}
\includegraphics[width= 0.85\textwidth]{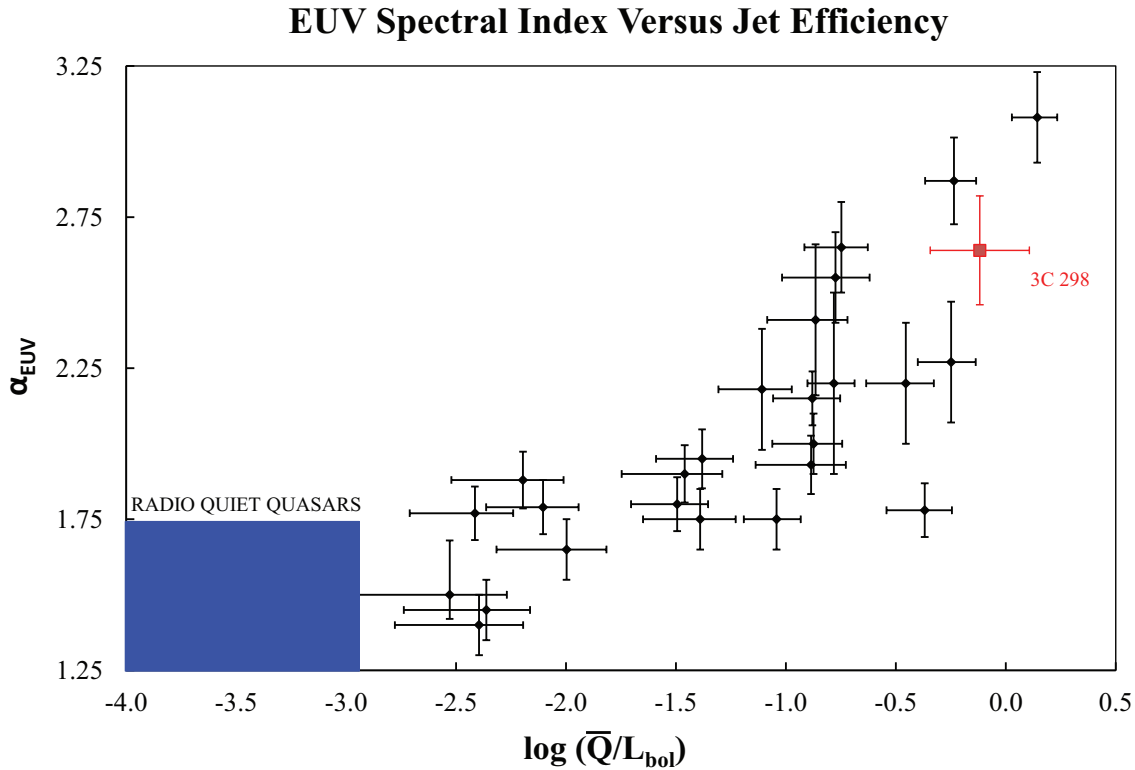}

\caption{3C 298 in the context of the EUV deficit of radio loud quasars relative to radio quiet quasars. The EUV deficit is quantified in terms of the spectral index in frequency space, $\alpha_{\rm{EUV}}$, taken from 1100~\AA\, to 700~\AA\, except in the case of 3C 298 where it is from 1100~\AA\, to 631~\AA\, (the center frequency of the GALEX far UV band in the quasar rest frame). The jet power normalized to the thermal luminosity from accretion is expressed as $\overline{Q}/L_{\rm{bol}}$.}
\end{center}
\end{figure}
Figure 6 reveals that the steepness of the SED in the EUV is not that unexpected based on the ratio of $\overline{Q}/L_{\rm{bol}}$ derived from Equations (6) and (14). We have added 3C 298 to the figure that appeared in \citet{pun21}. The correlated trend of $\alpha_{\rm{EUV}}$ with $\overline{Q}/L_{\rm{bol}}$ is known as the ``EUV deficit" of powerful radio loud quasars relative to radio quiet quasars. In the lower left hand corner, the objects with a ``small" value of $\overline{Q}/L_{\rm{bol}}$ have EUV continua indistinguishable from that of radio quiet quasars \citep{pun15}. This result is expected as for these quasars the jet output is an insignificant fraction of the total energy output of the quasar central engine. 3C 273 is an example of such a quasar.  The other points in Figure 6 are spectral indices from 1100~\AA\, to 700~\AA\, (quasar rest frame) derived from spectral data. For 3C 298 we computed $\alpha_{\rm{EUV}}$ from 1100~\AA\, to 631~\AA\, (the center frequency of the GALEX far UV band in the quasar rest frame) and we augmented the HST data with the GALEX data point after vetting its integrity in Section 7.1. The physical origin of the trend was explained by the quantitative consistency of the trend in Figure 6 with a scenario in which the jet is launched by large scale vertical magnetic flux in the inner region of the accretion flow, thereby displacing a proportionate fraction of the highest temperature thermal gas in the innermost accretion flow, i.e., the EUV emitting gas \citep{pun15}.
\par We proceed to explore the role of the weaker EUV continuum in producing the depressed EWs for the BELs. First consider the average $\alpha_{\rm{EUV}}=1.76$ for the HST observed quasars in the comparison sample of Table 3 \citep{tel02}. If we compute the ionizing luminosity, $L_{\rm{ion}}(\rm{HST})$ from 1 Rydberg to 10 Rydberg for the HST sample and calculate the same measure of the ionizing luminosity for 3C 298 based on Equation (13), $L_{\rm{ion}}(\rm{3C298})/L_{\rm{ion}}(\rm{HST}) =0.47$. The BEL flux, $F(\rm{BEL})$, does not necessarily scale linearly with the ionizing continuum, but has been described by a ``responsivity," $\eta$ \citep{kor04}.
\begin{equation}
F(\rm{BEL})\propto \Phi_{\rm{H}}^{\eta}\;, \Phi_{\rm{H}} \approx \int^{10\,\rm{Ryd}}_{1\,\rm{Ryd}}{[F_{\nu}(\rm{EUV})/\rm{h}\nu]\,d\nu};.
\end{equation}
where $\Phi_{\rm{H}}$ is the flux of hydrogen ionizing photons, and $F_{\nu}(\rm{EUV})$ is the EUV flux power law. We estimate that
\begin{equation}
\frac{\Phi_{\rm{H}}(\rm{3C298})}{\Phi_{\rm{H}}(\rm{HST})}=0.58\;.
\end{equation}
However, we do not have a robust estimate for $\eta$ for each BEL in a general circumstance. This issue is magnified in 3C 298 and other powerful radio loud quasars due to the large uncertainty in the coronal X-ray luminosity.

\section{Virial Black Hole Mass Estimates}
We start by choosing two virial black hole mass estimators that are based on MgII and L(3000) from the literature. The
formula from \citet{she12}, using the MgII BEL, $L(3000)= 5.34\times 10^{46}$ erg~sec$^{-1}$ and the FWHM of the total line profile (not tabulated in Table 2) of 3595 km~s$^{-1}$, yields
\begin{eqnarray}
&&\log\left(\frac{M_{bh}}{M_{\odot}}\right) = 1.815
+0.584\log\left(\frac{L(3000)}{10^{44} \,\rm{erg \ s}^{-1}} \right) +
1.712\log\left(\frac{\rm{FWHM}}{\rm{km \ s}^{-1}} \right)\;,\nonumber \\
&& \frac{M_{bh}}{M_{\odot}} = 3.13 \times 10^{9}  \;.
\end{eqnarray}
Alternatively, the formula of \citet{tra12} yields a different
estimate
\begin{equation}
\frac{M_{bh}}{M_{\odot}} = 5.6\times 10^{6}
\left(\frac{L(3000)}{10^{44}
\,\rm{erg}\, \rm{s}^{-1}}\right)^{0.5}\left(\frac{\rm{FWHM}}{1000
\,\rm{km \ s}^{-1}}\right)^{2} = 3.55 \times 10^{9}\;.
\end{equation}
\par These equations are predicated on the assumption that in general L(3000) is a uniform surrogate for the ionizing continuum, $L_{\rm{ion}}$. From Figure 5 and Section 8, however, we know that L(3000) is much larger than expected by direct comparison with the ionizing continuum and also indirectly from the EWs of the BELs in Table 3, suggesting that Equations (17) and (18) should over estimate $M_{bh}$. It is therefore probably more reasonable for this particular SED to use the line luminosity L(MgII) as a surrogate for $L_{\rm{ion}}$. We choose two estimators based on L(MgII) from the same two references as more accurate for this circumstance
The expression form \citet{she12} is
\begin{eqnarray}
&&\log\left(\frac{M_{bh}}{M_{\odot}}\right) = 3.979
+0.698\log\left(\frac{L({\rm MgII})}{10^{44} \,\rm{erg \ s}^{-1}}\right) +
1.382\log\left(\frac{\rm{FWHM}}{\rm{km \ s}^{-1}}\right)\;,\nonumber \\
&& \frac{M_{bh}}{M_{\odot}}= 2.36 \times 10^{9}  \;.
\end{eqnarray}
The analogous formula of \citet{tra12} yields a different
estimate
\begin{equation}
\frac{M_{bh}}{M_{\odot}} = 6.79\times 10^{6}
\left(\frac{L({\rm MgII})}{10^{42}
\,\rm{erg \ s}^{-1}}\right)^{0.5}\left(\frac{\rm{FWHM}}{1000
\,\rm{km \ s}^{-1}}\right)^{2} = 1.93 \times 10^{9}\;.
\end{equation}
\par Based on our analysis of the continuum near 3000~\AA\, (see Appendix B), the SED in Figure 5, and the EUV deficit described in Section 8, the best virial estimate of the central black hole mass is $M_{bh} = 2.14 \times 10^{9} M_{\odot}$ based on averaging Equations (19) and (20). The Eddington luminosity of such objects is $L_{\rm{Edd}}= 2.7\times 10^{47}$ erg~s$^{-1}$. Using the $L_{\rm{bol}}$ that was calculated in Section 7 produces a high Eddington ratio of 0.57. Alternatively, if we average Equations (17)-(20), $M_{bh} = 2.74 \times 10^{9}  M_{\odot}$ and $L_{\rm{bol}}/L_{\rm{Edd}}=0.45$. Both estimates are consistent with the rather large BLUE components found in CIV and Ly$\alpha$ in Table 2. Such components are thought to arise from radiation driven winds \citep{mur95,fin10}.

\section{Summary and Conclusion}
In a previous study it was noted that the quasars, 3C 82 and 3C 298, both members of the CSS class of radio sources, were candidates to have the largest known $\overline{Q}$ for a quasar \citep{pun20}. Hence, we initiated an equally detailed study of 3C 298 in this paper. In Section 2, we utilized the impressive new LOFAR observations of the radio source at 29-78 MHz. In Sections 2-5, we developed physical models of the lobe plasmas that indicate a long-term time-averaged jet power of, $\overline{Q} \approx 1.28 \pm 0.51 \times 10^{47} \rm{erg}\, \rm{s}^{-1}$ in Equation (6). This result is quantitatively similar, but smaller than the cruder standard estimate that is computed using only the 151 MHz flux density, $\overline{Q} \approx 2.18 \pm 0.71 \times 10^{47} \rm{erg \ s}^{-1}$ in Equation (8), justifying the need for a more detailed analysis for CSS sources.
\par Section 6 presented the rest frame SED of 3C 298 from 1100~\AA\, to 4300~\AA\, using two spectra from HET and an archival HST spectrum. Three component fits to the broad emission lines are given in Tables 2 and 3. The line strength ratios were typical of quasars, but the lines were relatively weak compared to the local continuum. To investigate further, we constructed an SED of just the continuum in Section 7 that was augmented with 2MASS and GALEX photometric points and \emph{Chandra} data. We deduced three extreme circumstances
\begin{enumerate}
\item 3C 298 potentially belongs to the rare class of kinetically dominated quasar jets, $\overline{Q}/L_{\rm{bol}}\approx 0.87 \pm 0.41$ (see Figure 6), similar to the other potentially kinetically dominated sources tabulated in \citet{pun20}, $\overline{Q}/L_{\rm{bol}}\approx 1$.
\item The EUV continuum was extremely steep, $\alpha_{\rm{EUV}} = 2.64 \pm 0.18$ compared to  $\alpha_{\rm{EUV}}=1.76\pm 0.12$ for the HST observed quasars of comparable luminosity \citep{tel02}.
\item The cosmological rest frame 1.2-24 keV, X-ray luminosity of $5.17\times 10^{46}$ erg~s$^{-1}$ is extremely large for a lobe-dominated steep spectrum radio source (a non-blazar).
\end{enumerate}
In Section 8, we tried to reconcile the small BEL EWs with the steep EUV continuum. It is highly plausible that the weak ionizing EUV continuum is the reason for the depressed lines strengths. Even though there is poor coverage of the EUV spectrum in Figure 5, Figure 6 demonstrates that our data place 3C298 in good agreement with the EUV deficit trend that is seen in powerful lobe dominated quasars. The interpretation of the weak BELs and the EUV deficit seems reasonable.
\par Section 9 considers the relatively weak EUV in the context of virial mass estimates that utilize near UV luminosity. We argue that in this case it is more consistent to use the BELs themselves as a surrogate for $L_{\rm{ion}}$, and determined that the Eddington rate of accretion is $\approx 0.5$. This high Eddington rate is an indication of a powerful source of radiation pressure.  This large accretion is imprinted in the BELs as strong BLUE components in Table 2 and Figure 8 (Appendix B) in Ly$\alpha$ and CIV. Evidence of powerful baryonic outflows is commonly seen in both UV absorption and UV emission in high Eddington rate radio quiet quasars, but is rare in radio-loud quasars with an FRII morphology \citep{ric02,pun10,bec00,bec01}. Other 3C CSS quasars, such as 3C82 and 3C 286, also have this unusual property, for a radio-loud quasar, of excess blue-shifted CIV emission \citep{pun20}. We have obtained deep HET optical spectra of 3C CSS quasars to augment the HST archives, and plan to perform a complete spectral analysis and report our findings in a future work.
\par This work motivates further investigations of the powerful X-ray jet in this mildly Doppler enhanced source. Figure 2 and Table 1 suggest the possibility of some modest variability of the nuclear region above 5 GHz (as discussed in Sections 4 and 5), so there might be some modest relativistic behavior. In spite of point 3 above, 3C298 is not a gamma ray source. It is, however, quite bright, and would be an excellent target for NuSTAR to determine if there is a turnover above 100 keV in the cosmological rest frame.

\begin{acknowledgments}
CG acknowledges support from the ERC Starting Grant ClusterWeb 804208. We thank the staff of the Hobby Eberly Telescope for support of our VIRUS and LRS2 observations. This research was partially funded through the McDonald Observatory.

The Hobby-Eberly Telescope (HET) is a joint project of the University of Texas at Austin, the Pennsylvania State University, Ludwig-Maximilians-Universität München, and Georg-August-Universität Göttingen. The HET is named in honor of its principal benefactors, William P. Hobby and Robert E. Eberly.

VIRUS is a joint project of the University of Texas at Austin, Leibniz-Institut f{\" u}r Astrophysik Potsdam (AIP), Texas A\&M University (TAMU), Max-Planck-Institut f{\" u}r Extraterrestrische Physik (MPE), Ludwig-Maximilians-Universit{\" a}t M{\" u}nchen, Pennsylvania State University, Institut f{\" u}r Astrophysik G{\" o}ttingen, University of Oxford, and the Max-Planck-Institut f{\" u}r Astrophysik (MPA).

The Low Resolution Spectrograph 2 (LRS2) was developed and funded by the University of Texas at Austin McDonald Observatory and Department of Astronomy and by Pennsylvania State University. We thank the Leibniz-Institut f{\" u}r Astrophysik Potsdam (AIP) and the Institut f{\" u}r Astrophysik G{\" o}ttingen (IAG) for their contributions to the construction of the integral field units.

We acknowledge the Texas Advanced Computing Center (TACC) at The University of Texas at Austin for providing high performance computing, visualization, and storage resources that have contributed to the results reported within this paper.

This work makes use of the Pan-STARRS1 Surveys (PS1) and the PS1 public science archive, which have been made possible through contributions by the Institute for Astronomy, the University of Hawaii, the Pan-STARRS Project Office, the Max-Planck Society and its participating institutes.
This work makes use of data from the European Space Agency (ESA) mission {\it Gaia} (https://www.cosmos.esa.int/gaia), processed by the {\it Gaia} Data Processing and Analysis Consortium (DPAC, https://www.cosmos.esa.int/web/gaia/dpac/consortium). Funding for the DPAC has been provided by national institutions, in particular the institutions participating in the {\it Gaia} Multilateral Agreement.
This work makes use of the Sloan Digital Sky Survey IV, with funding provided by the Alfred P. Sloan Foundation, the U.S. Department of Energy Office of Science, and the Participating Institutions. SDSS-IV acknowledges support and resources from the Center for High-Performance Computing at the University of Utah. The SDSS web site is www.sdss.org.
\end{acknowledgments}

\appendix
\section{The Underlying Physical Equations}
To allow this article to be self-contained, this brief appendix repeats the mathematical formalism associated with spherical plasmoids previously described in
detail in \citet{pun20}. First, one must differentiate between quantities measured in the plasmoid frame of reference and those measured in the observer's frame of reference.
The physics is evaluated in the plasma rest frame. The results are then transformed to the observer's frame. The underlying power law for the
flux density is defined as $S_{\nu}(\nu_{\mathrm{obs}}) = S\nu_{\mathrm{obs}}^{-\alpha}$, where $S$ is a constant. Observed quantities will
be designated with a subscript, ``obs'', in the following expressions. The observed frequency is related to the emitted frequency, $\nu_{\mathrm{e}}$, by $\nu_{\mathrm{obs}} =\delta \nu_{\mathrm{e}}/(1+z)$, where $\delta$ is the bulk flow Doppler factor, $\delta \approx 1$. The SSA attenuation coefficient is computed in the plasma rest frame \citep{gin69},
\begin{eqnarray}
&& \mu(\nu_{\mathrm{e}})=\overline{g(n)}\frac{e^{3}}{2\pi
m_{e}}N_{\Gamma}(m_{e}c^{2})^{2\alpha} \left(\frac{3e}{2\pi
m_{e}^{3} c^{5}}\right)^{\frac{1+2\alpha}{2}}\left(B\right)^{(1.5
+\alpha)}\left(\nu_{\mathrm{e}}\right)^{-(2.5 + \alpha)}\;,\\
&& \overline{g(n)}= \frac{\sqrt{3\pi}}{8}\frac{\overline{\Gamma}[(3n
+ 22)/12]\overline{\Gamma}[(3n + 2)/12]\overline{\Gamma}[(n +
6)/4]}{\overline{\Gamma}[(n + 8)/4]}\;, \\
&& N=\int_{\Gamma_{\mathrm{min}}}^{\Gamma_{\mathrm{max}}}{N_{\Gamma}\Gamma^{-n}\,
d\Gamma}\;,\; n= 2\alpha +1 \;,
\label{equ:A2}
\end{eqnarray}
where $\Gamma$ is the ratio of lepton energy to rest mass energy, $m_{e}c^2$, $\overline{g(n)}$ is the Gaunt factor averaged over angle, $\overline{\Gamma}$ is the gamma
function, and $B$ is the magnitude of the total magnetic field. The low energy cutoff is $E_{\mathrm{min}} = \Gamma_{\mathrm{min}}m_{e}c^2$.

A simple solution to the radiative transfer equation occurs in the homogeneous approximation \citep{gin65,van66}
\begin{eqnarray}
&& S_{\nu_{\rm{obs}}} = \frac{S_{\rm{o}}\nu_{\rm{obs}}^{-\alpha}}{\tau(\nu_{\rm{obs}})} \times \left(1 -e^{-\tau(\nu_{\rm{obs}})}\right)\;, \; \tau(\nu_{\rm{obs}})
\equiv \mu(\nu_{\rm{obs}}) L\;, \; \tau(\nu_{\rm{obs}})=\overline{\tau}\nu_{\rm{obs}}^{(-2.5 +\alpha)}\;,
\label{equ:A5}
\end{eqnarray}
where $\tau(\nu)$ is the SSA opacity, $L$ is the path length in the rest frame of the plasma, $S_{\rm{o}}$ is a normalization factor and $\overline{\tau}$ is a constant.

Connecting the parametric spectrum given by Equation~(\ref{equ:A5}) to a physical model requires an expression for the synchrotron emissivity \citep{tuc75}:
\begin{eqnarray}
&& j_{\nu} = 1.7 \times 10^{-21} [4 \pi N_{\Gamma}]a(n)B^{(1
+\alpha)}\left(\frac{4
\times 10^{6}}{\nu_{e}}\right)^{\alpha}\;,\\
&& a(n)=\frac{\left(2^{\frac{n-1}{2}}\sqrt{3}\right)
\overline{\Gamma}\left(\frac{3n-1}{12}\right)\overline{\Gamma}\left(\frac{3n+19}{12}\right)
\overline{\Gamma}\left(\frac{n+5}{4}\right)}
       {8\sqrt\pi(n+1)\overline{\Gamma}\left(\frac{n+7}{4}\right)} \;,
\label{equ:A6}
\end{eqnarray}
where the coefficient $a(n)$ separates the pure dependence on $n$ \citep{gin65}.
One can transform this result to the observed flux density, $S(\nu_{\rm{obs}})$, in the optically thin region of the spectrum using the relativistic transformation relations from
\citet{lin85},
\begin{eqnarray}
 && S(\nu_{\rm{obs}}) = \frac{\delta^{(3 + \alpha)}}{4\pi D_{L}^{2}}\int{j_{\nu}^{'} d V{'}}\;,
\label{equ:A8}
\end{eqnarray}
where $D_{L}$ is the luminosity distance and in this expression, the
primed frame is the rest frame of the plasma.

\subsection{Mechanical Quantities that Characterize the Lobes} First define the kinetic energy of the protons, $KE(\mathrm{proton})$,
\begin{eqnarray}
 && KE(\mathrm{protonic}) = (\gamma - 1)Mc^{2}\;,
\end{eqnarray}
here $M$ is the mass of the plasmoid. Secondly define the lepto-magnetic energy, $E(\mathrm{lm})$ which creates the synchrotron emission. It is the volume integral of the leptonic internal energy density, $U_{e}$, and the magnetic field energy density, $U_{B}$. $E(\mathrm{lm})$ in a uniform spherical volume is
\begin{eqnarray}
 && E(\mathrm{lm}) = \int{(U_{B}+ U_{e})}\, dV = \frac{4}{3}\pi R^{3}\left[\frac{B^{2}}{8\pi}
+ \int_{\Gamma_{min}}^{\Gamma_{max}}(m_{e}c^{2})(N_{\Gamma}E^{-n + 1})\, d\,E \right]\;.
\end{eqnarray}
The leptons also have a kinetic energy analogous to Equation (11),
\begin{eqnarray}
 && KE(\mathrm{leptonic}) = (\gamma - 1)\mathcal{N}_{e}m_{e}c^{2}\;,
\end{eqnarray}
where $\mathcal{N}_{e}$ is the total number of leptons in the plasmoid.
\section{Fits to the Broad Line Profiles}
\begin{figure*}
\begin{center}
\includegraphics[width= 0.4\textwidth,angle =0]{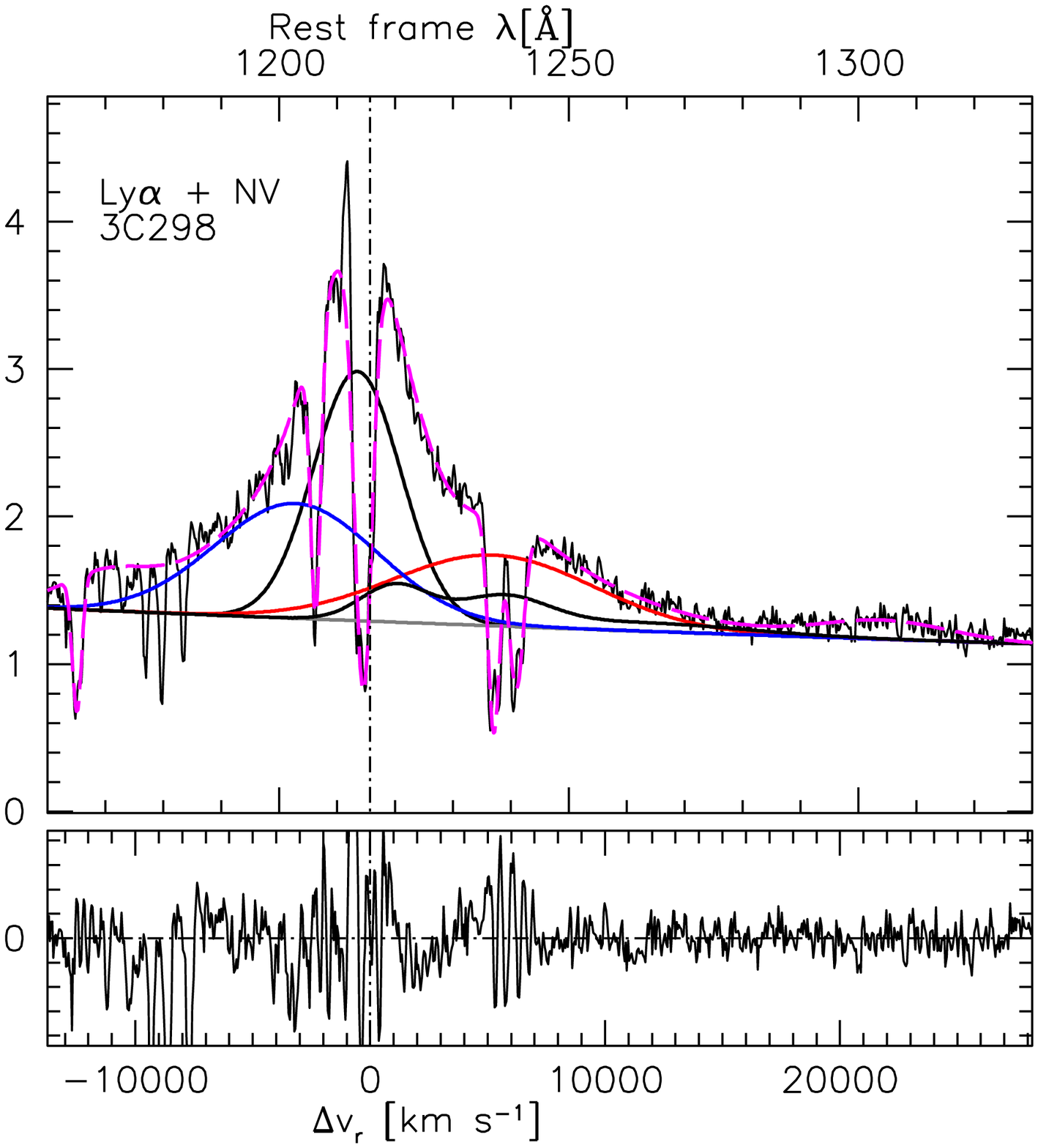}
\includegraphics[width= 0.4\textwidth,angle =0]{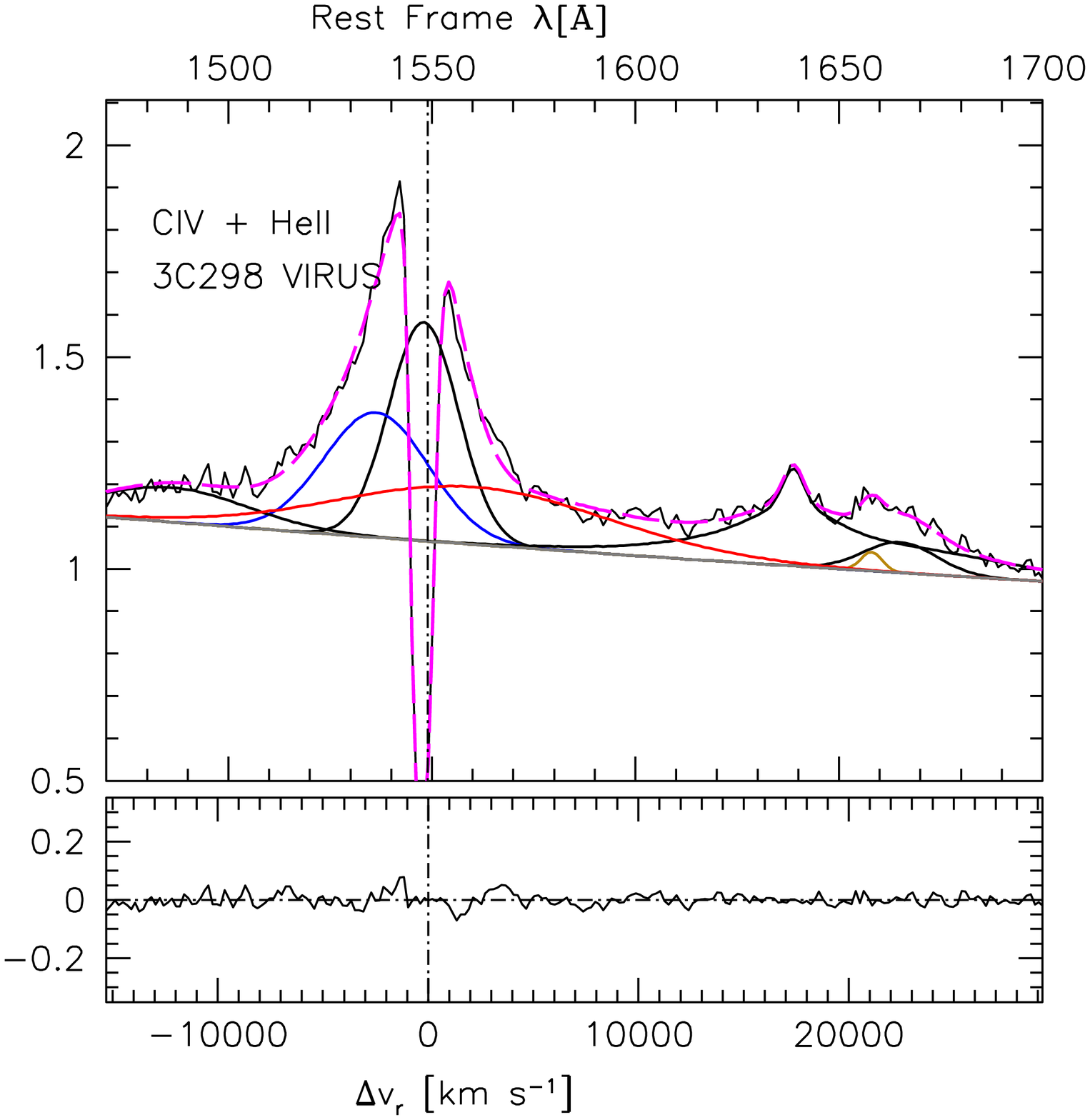}
\includegraphics[width= 0.4\textwidth,angle =0]{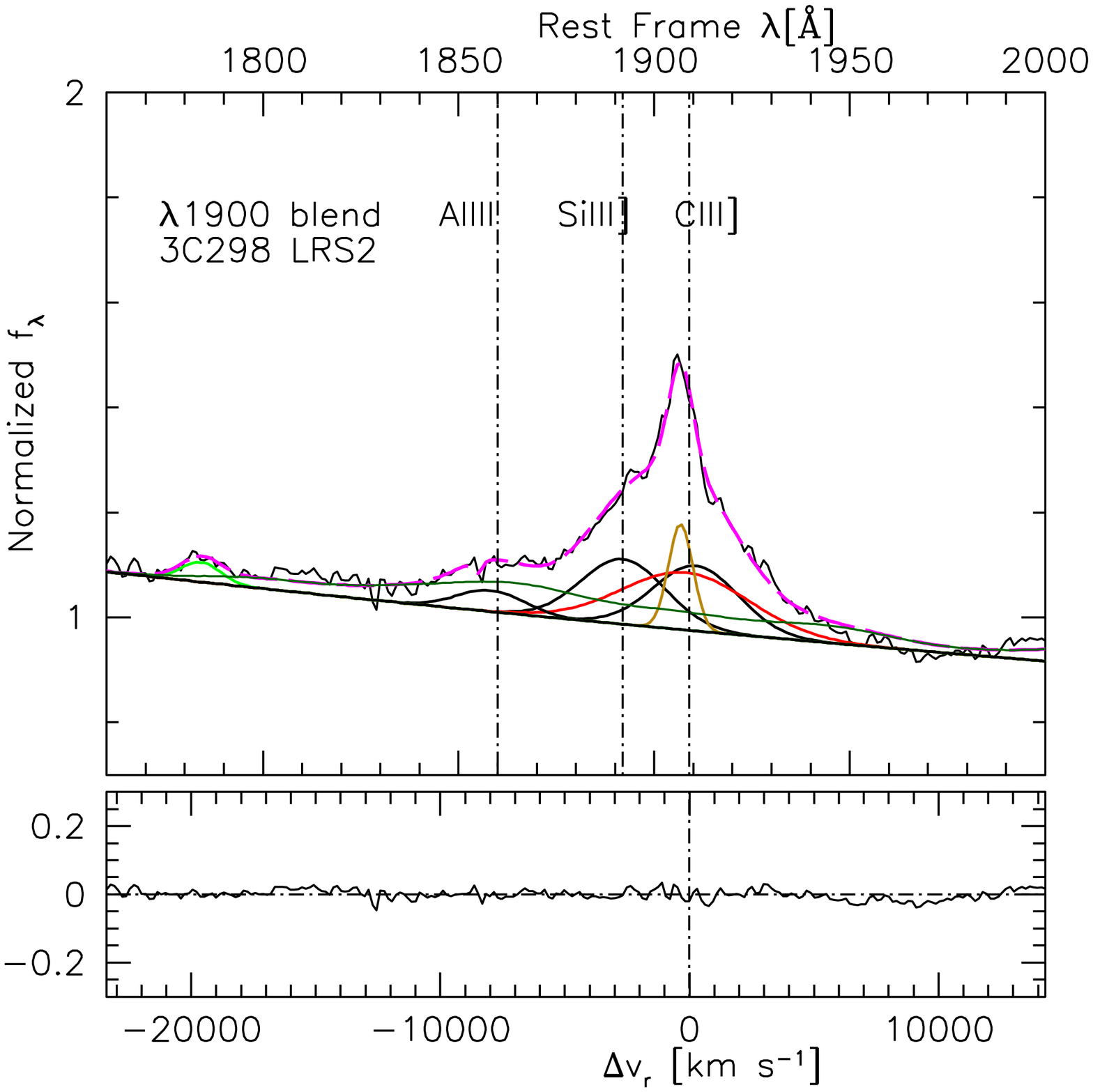}
\includegraphics[width= 0.55\textwidth, bb=0 210 600 700, clip=, angle=0]{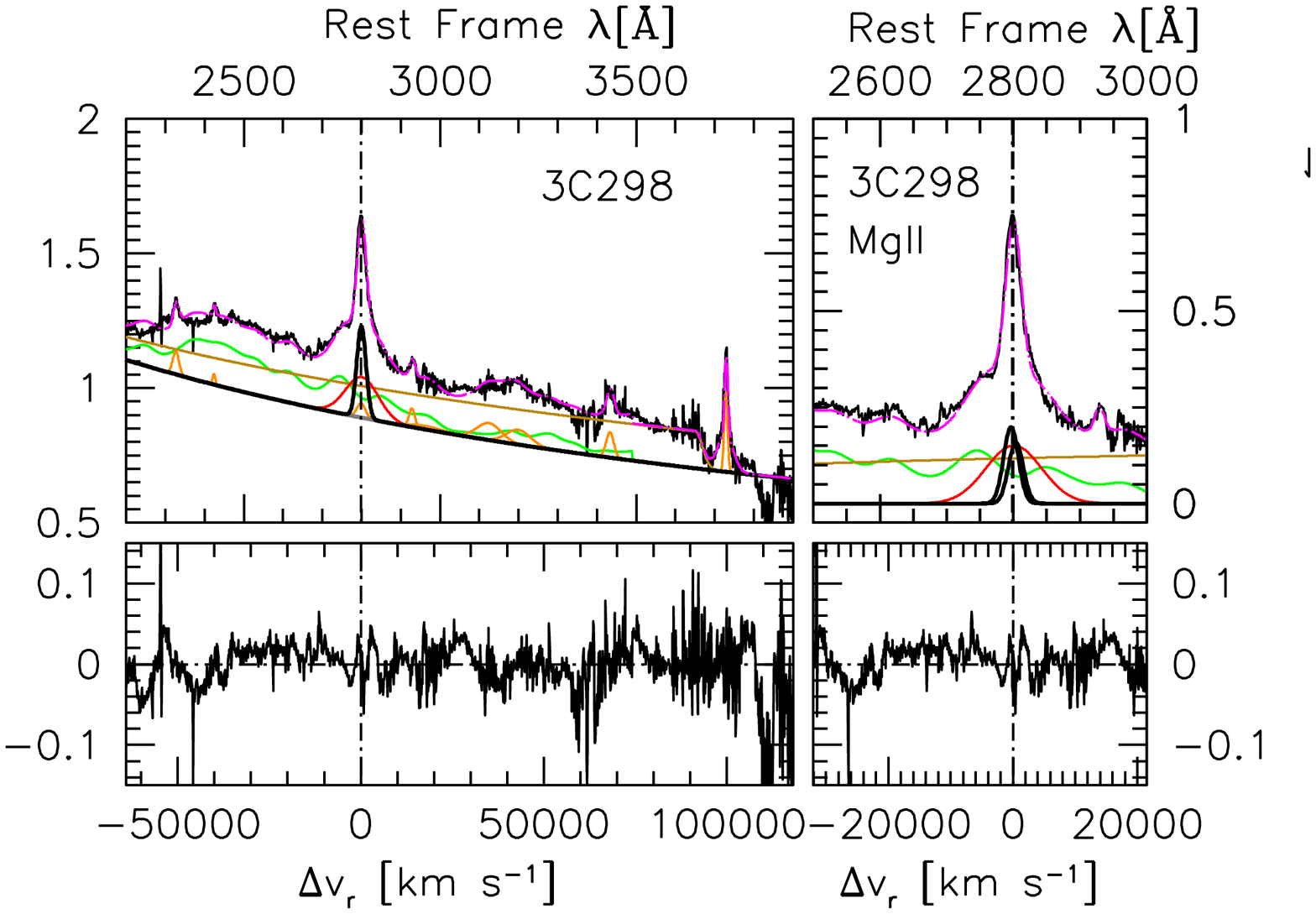}

\caption{\footnotesize{Line fits from Table 2 for the four regions of the spectrum with strong UV emission lines. The flux density is in relative units with a scaling of 1 being near the continuum level. The (black) broad line in the blue wing of CIV is NIV]$\lambda$1486. The MgII continuum is masked by a strong Balmer continuum, the brown curve, and FeII in green. The BLUE component in blue, the VBC is red and the BC is black. HeII and NIV are entirely black. In the lower right set of panels that highlight the MgII region, the FeII complex is in green and the Balmer continuum is brown.}}
\label{fig:BEL}
\end{center}
\end{figure*}

The BEL components are identified in Figure~\ref{fig:BEL} as follows: the BC is the black Gaussian profiles, the VBC is red curve and BLUE is the blue curve. Only the sum of the three components is shown for both NV$\lambda$1240 and HeII$\lambda$1640 (in black). For more details of the fitting process see \citep{pun20}.

Modelling of the Balmer continuum in the MgII region, in the lower right hand panel of Figure~\ref{fig:BEL}, is nontrivial. The model spectrum depends upon numerous free parameters, an intensity factor, electron temperature, $T_{e}$, and the optical depth \citep{gra82,jin12}. Since we lack adequate spectral coverage of the longer wavelength discrete Balmer lines (long-ward of 4000 \AA), there is little information to constrain the Balmer continuum. We simply assume an optically thin continuum and $T_{e} =17500\rm{K}$ \citep{kwa81,jin12}. Our primary interest is to derive the MgII broad line parameters and define the SED continuum level. In order to assess if our modeling  of the small blue bump at 3000~\AA\, is reasonable, we compare our results to the \citet{she11} fit of the SDSS spectrum. Our (their) $L(3000\AA\,) = 5.36 \times 10^{46}$ ergs $\rm{s}^{-1}$ ($L(3000\AA\,) = 4.60 \times 10^{46}$ ergs $\rm{s}^{-1}$ ), total BEL FWHM is 3595 km~s$^{-1}$ (3908 km~s$^{-1}$) and the broad line luminosity is $L$(MgII) is $4.60 \times 10^{44}$ erg~s$^{-1}$ ($3.72 \times 10^{44}$ erg~s$^{-1}$). The small differences between the two fits is likely attributable to a larger flux allocation to the UV FeII and the Balmer continuum than our fit. There was no effort in \citet{she11} in their UV FeII excision to assign an explicit contribution to a Balmer pseudo-continuum. Even though ours was a dedicated fit of an individual source, we cannot claim that our fit is better; the two are similar and produce bolometric luminosities that agree to within $\sim 1\%$.
\end{document}